\documentclass[twocolumn,showpacs,preprintnumbers,amsmath,amssymb,superscriptaddress,prc]{revtex4}
\usepackage{graphicx}
\usepackage{epstopdf}
\usepackage{dcolumn}
\usepackage{bm}
\usepackage[english]{babel}
\usepackage{amsmath,amssymb}
\usepackage{graphicx}
\usepackage{multirow}
\usepackage{color}
\usepackage[normalem]{ulem}
\usepackage{tikz}

\usepackage{float}
\restylefloat{figure}

\usepackage{mathrsfs}

\newcommand{\ket}[1]{ | #1 \rangle }

\newcommand{\overlap}[2]{\langle #1 | #2 \rangle}
\newcommand{\elmx}[3]{\langle #1 | #2 | #3 \rangle}

\begin{document}

\title{Nuclear Structure with Discrete Non-Orthogonal Shell Model: new frontiers}

\author{D. D. Dao}
\author{F.~Nowacki}
 \affiliation{Universit\'e de Strasbourg, CNRS, IPHC UMR7178, 23 rue du Loess, F-67000 Strasbourg, France}  
\date{\today}

\begin{abstract}
We present developments and applications for the diagonalization of shell-model hamiltonians in a discrete non-orthogonal basis (DNO-SM). The method, and its actual numerical implementation \texttt{CARINA}, based on mean-field and beyond-mean field techniques has already been applied in previous studies and is focused on basis states selection optimization. The  method is benchmarked against a full set of $sd$ shell exact diagonalizations, and is applied for the first time to the heavy deformed $^{254}$No nucleus.
\end{abstract}


\maketitle


\section{Introduction}
In the recent decades,
the advent of radioactive beam factories associated with developments of more sophisticated experimental methods has enabled to discover new manisfestations of many-body nuclear dynamics in many places of the nuclear chart. New phenomena like halo systems, two-proton radioactivity, occurence of new magic numbers, vanishing of shell closures, soft dipole modes, or even searches for superheavy nuclei
have been observed and have stimulated the continuous developments and improvements of theoretical methods in order to interpret such phenomena. 

Among the various theoretical frameworks available, the Shell-Model (SM) or Configuration Interaction (CI), either in its no-core or valence space implementations,  has always been one of the most powerful methods in the description of quantum nuclear systems~\cite{RMP,PPNP}, in particular with the numerical development of efficient diagonalization codes which have opened the era of the so-called "Large Scale Shell-Model calculations" for light and medium-mass nuclei up to $A\sim 150$~\cite{Caurier:1998zw,Nowacki:2016isq,Sn100,Siciliano:2019qhw,MINIBALL:2018zvw,Naidja:2017tyv} and have become the method of choice to explain the observed nuclear phenomena, guide experimental programs, and not the least, allowed for a deeper understanding of man astrophysical objects and processes in which exotic nuclei often play the key role. 
However, its success was always minored by the exponential growth 
of the systems basis involved. 
At the same time, variational methods with symmetry breaking and restoration have also shown great success for decades and have proven to a certain extent to be applicable over the nuclear chart from the lightest nuclear systems to the most heavy ones~\cite{Bender:2003jk,Egido:2016bdz,Robledo:2018cdj}. 

Although these methods provide distinct description,
the merging of the mean-field techniques within the shell-model formalism has already been developed and studied in the literature in the past, starting from the pioneering work of Ripka~\cite{ripk1,Ripka1968}, later followed by the different \texttt{VAMPIR} implementations~\cite{VampirI,VampirII}. One of the major achievements up to now was proposed by the Tokyo group with the Monte Carlo Shell-Model~\cite{MCSM,MCSMII,MCSMIII}. And more recently several implementations were used, either in an punctual manner~\cite{ ,zch09,hke11} or in a more ambitious scale with the recent development of the \texttt{TAURUS} numerical suite~\cite{Taurus,Taurus-sd,Taurus-pf}.

In the present work, we present the formalism of the Discrete Non-Orthogonal Shell-Model (DNO-SM)
and its associated numerical implementation \texttt{CARINA}. The DNO-SM amounts to diagonalize valence shell-model hamiltonians in a non-orthogonal basis with the use of beyond-mean-field techniques. 
The detailled framework is exposed in the next section and applications to $sd$ shell nuclei in comparison with exact diagonalisations are discussed in section~\ref{Benchmark and comparison in sd nuclei}. The final section exposes an application to a very heavy system in the $^{254}$No case. 

\section{Theoretical Framework}
\subsection{Shell Model formulation revisited}
\subsubsection{The diagonalization dilemma}

The ultimate question in the Shell Model, once a physically meaningful valence space $\mathcal E$ is equipped together with the associated effective interaction $\hat{\mathcal V}$ for the problem at hand, is to tackle the secular equation
\begin{equation}
  \hat{\mathcal H} \ket\Psi = E\ket\Psi
  \label{eq:secular_equation}
\end{equation}
where $\hat{\mathcal H}$ represents the effective Hamiltonian composed of $\hat{\mathcal V}$ and a one-body single-particle energy $\{e_i\}$ part
\begin{equation}
  \hat{\mathcal H} = \sum_{i\in\mathcal E} e_i a^\dagger_ia_i + \frac{1}{4}\sum_{ijkl\in\mathcal E}
  \elmx{ij}{\hat{\mathcal V}}{kl} a^\dagger_ia^\dagger_ja_la_k.
  \label{eq:hamiltonian}
\end{equation}
$\{a^\dagger_i\}$ and $\{a_i\}$ are creation and annihilation operators satisfying the common anti-commutation rules for fermionic systems.

By defining a set of basis states $\mathcal B=\{\ket{\phi_m},m\in\mathbb N\}$ constructed from the single-particle spherical oscillator valence space $\mathcal E$ with which we can write the eigenstate $\ket\Psi$ as
\begin{equation}
  \ket\Psi = \sum_{m=1}^{\mathrm{dim}(\mathcal B)} c_m\ket{\phi_m},
\end{equation}
the classic Shell Model resolution of~\eqref{eq:secular_equation} then amounts to addressing the eigenvalue problem
\begin{equation}
  \sum_{m=1}^{\mathrm{dim}(\mathcal B)} \mathcal H_{m'm}c_m = E\,c_{m'}
  \label{eq:projected_secular_equation}
\end{equation}
by an exact diagonalization of the Hamiltonian matrix $\mathcal H_{m'm} = \elmx{\phi_{m'}}{\hat{\mathcal H}}{\phi_m}$ in the model space $\mathscr H = \overline{\mathrm{Span}\:\mathcal B}$. Although spectacular designs of Shell Model codes have been achieved~\cite{RMP} to reach space dimensions larger and larger, it is still an inherent problem that makes difficult to extend the Shell Model applicability into heavier mass nuclei. To deal with this dilemma, we look for a replacement of $\mathcal B$ with a different family of basis states. The existence of such basis takes the root in the original idea of the Generator Coordinate Method (GCM) first proposed in Refs.~\cite{GCM1957a,GCM1957b}. As we shall discuss in the following, it can be however viewed in an independent status with respect to the GCM, thanks to the work of the authors in Ref.~\cite{Deumens1979}.

\subsubsection{Discrete non-orthogonal basis}

The starting point of the GCM is the hypothesis that one can find a family of states depending on the continuous (generator) coordinate(s) $q$
\begin{equation}
  \mathit\Gamma = \{\ket{\Phi(q)}\:|\: q\in\mathbb R\}
\end{equation}
so that the latter forms a model subspace $\mathscr{H}_q = \overline{\mathrm{span}\:\mathit\Gamma} \subseteq \mathscr H$ following the nature of the coordinate(s) $q$ that we choose in the generation of $\mathit\Gamma$. The core of our presentation of $\mathscr H_q$ relies on the following existence theorem first noticed in Ref.~\cite{Deumens1979}. Suppose $\mathscr H_q$ is a separable Hilbert subspace, i.e. $\mathscr H_q \subseteq \mathscr H \subset \mathscr{L}^2$ where $\mathscr{L}^2$ denotes the full Hilbert space associated with the space of square integrable functions, the separability property of $\mathscr H_q$ implies the existence of a countable family
\begin{equation}
  \mathit\Gamma_0 = \{\ket{\Phi(q_i)}\:|\:i\in\mathbb N\} \subset \mathit\Gamma
  \label{eq:DNObasis}
\end{equation}
which is in general skew or non-orthogonal set of states with the property $\mathscr H_q = \overline{\mathrm{span}\:\mathit\Gamma_0}$ (cf. the detailed demonstration given in Appendix of Ref.~\cite{Deumens1979}). This enables us to tackle now the diagonalization of $\hat{\mathcal H}$ in the subspace $\mathscr H_q$ represented by the discrete non-orthogonal basis set $\mathit\Gamma_0$. Indeed, by expressing the eigenstate $\ket\Psi$ as
\begin{equation}
  \ket\Psi = \sum_{i=0}^{\infty} f(q_i) \: \ket{\Phi(q_i)},
  \label{eq:non_orthogonal_expansion}
\end{equation}
the projection of~\eqref{eq:secular_equation} in $\mathscr H_q$ becomes equivalent to the generalized eigenvalue problem
\begin{equation}
  \sum_{i=0}^{\infty}\Big[\mathcal H(q_{i'},q_i) - E\;\mathcal N(q_{i'},q_i)\Big] f(q_i) = 0
  \label{eq:projected_Hq_equation}
\end{equation}
where $\mathcal O(q_{i'},q_i) = \elmx{\Phi(q_{i'})}{\hat{\mathcal O}}{\Phi(q_i)}$ ($\hat{\mathcal O} = \hat{\mathcal H}, \mathbf 1$) are the Hamiltonian and norm matrix elements and $f(q_i)$ the expansion coefficient.

Therefore, instead of using the basis $\mathcal B$ spanning the full model space $\mathscr H$, the above presented theorem on the existence (not necessarily unique) of a discrete non-orthogonal basis set of $\hat{\mathcal H}$ suggests that:
\begin{enumerate}
\item the diagonalization of $\hat{\mathcal H}$ in $\mathscr H_q$ becomes ``exact'' when $\mathscr H_q = \mathscr H$, which means the coordinate(s) $q$ must be chosen so as to ``exhaust'' in some way the space $\mathcal E$;
\item the truncation of the infinite countable set $\mathit\Gamma_0$ could be done in a variational way such that
the finite sum
\begin{equation}
  \ket\Psi \approx \sum_{i=0}^n f(q_i) \ket{\Phi(q_i)}
\end{equation}
yields an optimal approximation.
\end{enumerate}
Whether we are able to choose $q$ to fulfil the condition $\mathscr H_q = \mathscr H$ can be verified a posteriori. What needed is then an efficient truncation method of $\mathit\Gamma_0$, which we shall address now.

\subsubsection{Truncation method with the minimization technique}

The existence theorem as presented previously has enabled us to transform the classic Shell Model eigenvalue problem~\eqref{eq:projected_secular_equation}, formulated in the orthonormal basis $\mathcal B$, into the generalized one~\eqref{eq:projected_Hq_equation} through the discrete non-orthogonal basis $\mathit\Gamma_0$. As noted earlier by the authors of Ref.~\cite{Deumens1979}, the minimization technique originally proposed by E. Caurier in Ref.~\cite{Caurier1975} provides an iterative prescription to truncate $\mathit\Gamma_0$ in a variational way.

It proceeds as follows \cite{Caurier1975}: \textit{``the first point $q_0$ is the one such that $\ket{\Phi(q_0)}$ minimizes the energy. The second point $q_1$ is chosen in such a way that the energy obtained from diagonalizing the Hamiltonian in the $2$--dimensional space spanned by $\ket{\Phi(q_0)}$ and $\ket{\Phi(q_1)}$ be a minimum. One proceeds in the same way to determine the third basis vector $\ket{\Phi(q_2)}$ etc...''}.

The technique was however implemented only in toy model examples of Hydrogen atom in molecular physics (cf. e.g.~\cite{Arickx1981}) and has never been considered in realistic nuclear structure calculations. Therefore, instead of the conventional problem represented by $(\hat{\mathcal H}, \mathscr H, \mathcal B)$, our current work exploits fully this technique for the first time in the Shell Model framework formulated in terms of $(\hat{\mathcal H}, \mathscr H_q, \mathit\Gamma_0)$. This resulting model will be from now on referred to as \textit{Discrete Non-Orthogonal Shell Model} (DNO-SM).

To be now more precise in the practical realization of the DNO-SM, we assume that the Projected Constrained Hartree--Fock (PCHF) approach provides us with a basis generation method which will be our focus in the next subsection. However, before going further, let us note that, the above proposed technique suggested to minimize one state at a time. More generally in order to cover broader physical situations such as nuclear coexistences where the potential energy surface may exhibit several local minima or when states are not of collective nature, it may be more preferable to obtain various excited states by a single minimization. A generalization to deal with the minimization of many states simultaneously shall be presented.

\subsection{Projected constrained Hartree-Fock basis}

Having introduced the general framework of our approach to the dimensionality problem encountered in the classic Shell Model, we present now the construction of the many-body basis in the DNO-SM. The choice of degrees of freedom here is important to take into account correlations as much as possible in the generation of the basis. This should be inferred on physical grounds. Moreover, the many-body basis must conserve important symmetries of the effective Hamiltonian, in particular, the cases associated with conserved quantities such as the angular momentum and particle numbers. Such basis could be built upon the Constrained Hartree-Fock (CHF) method which relies on the rotational symmetry breaking at the mean field to incorporate deformations. A projection onto good angular momentum can be applied later before proceeding to the full diagonalization.

A Hartree-Fock (HF) state $\ket{\Phi(q)} = \displaystyle \prod_{i=1}^A a^\dagger_i\ket{0}$ for a nucleus of $A$--particles is obtained from CHF calculations under the conditions
\begin{align}
  \frac 1 2 \elmx{\Phi}{\hat Q_{\lambda\mu} + (-)^\mu \hat Q_{\lambda-\mu}}{\Phi} = Q_{\lambda\mu},\\
  \elmx{\Phi}{\hat J_m}{\Phi} = \langle\hat{J}_m\rangle \:\:(m = x,z),
  \label{eq:contraints}
\end{align}
where $\hat Q_{\lambda\mu} = r^{\lambda}Y_{\lambda\mu}(\theta,\varphi)$ is the multipole operator expressed in terms of the spherical harmonics $Y_{\lambda\mu}$ and $\hat J_m$ the components of the total angular momentum operator $\hat{\mathbf J}$. Here $Q_{\lambda\mu}$ and $\langle\hat{J}_m\rangle$ are desired constraining expectation values. Our numerical implementations of constrained HF calculations follow the standard modified Broyden method as described in Refs.~\cite{BroydenI,BroydenII,BroydenIII}. Furthermore, to ensure the correct constraining value along the iterative HF process, we use the augmented Lagrange method proposed in Ref.~\cite{augLag}.

Once this is done, the resulting HF state is projected onto good angular momentum $J$ through the usual procedure using the projection operator~\cite{projI,Ripka1968}
\begin{equation}
  \begin{aligned}
    &\mathcal P^{J}_{MK}(A) = \frac{2J+1}{4\pi^2\big(3-(-)^A\big)} \times \\
    &\int_0^{2\pi}d\alpha\int_0^{\pi}d\beta\int_{0}^{\gamma_{\rm max}}d\gamma \:
    \mathcal D^{J*}_{MK}(\alpha,\beta,\gamma) \: \hat R(\alpha,\beta,\gamma),
  \end{aligned}
\end{equation}
where $\gamma_{\rm max}=\big(3-(-)^A\big)\pi$, $\hat R(\alpha,\beta,\gamma)$ and $\mathcal D^{J*}_{MK}(\alpha,\beta,\gamma)$ denote the rotation operator and the Wigner matrix~\cite{WignerD}. The case of $\gamma_{\rm max} = 4\pi$ corresponds to a HF state describing odd systems. Thus we have included the dependence of the projection operator on the mass number $A$ for convenience. This provides us a family of PCHF states characterized by the angular momentum projection onto the intrinsic axis $|K|\leq J$ and the coordinate $q$ for a given $J$
\begin{equation}
  \mathit\Gamma = \{\mathcal P^J_{MK}(A)\ket{\Phi(q)}\:|\: q\in\mathbb R\}
\end{equation}
with which we can now formulate the DNO-SM's working equations.

The HF procedure to generate deformed Slater determinants in the construction of DNO-SM basis is implemented to treat both odd-- and even--nuclei without further assumptions. More specifically, we do not impose any self--consistent symmetries (e.g. no time-reversal, no parity conservation) at the HF mean field to exploit at best what is offered by the single-particle valence space $\mathcal E$. The construction of DNO-SM basis for odd nuclei is done via the constrain of angular momentum components $\hat J_m\:(m=x,z)$ which, as we will show later, can provide a very good many-body basis for such nuclei. The formal aspect of this approach to treat odd-mass nuclei has been pointed out in Ref.~\cite{crankingI}. More recently, the theoretical demonstration in the Hartree-Fock-Bogoliubov framework is described in Ref.~\cite{crankingII}.

Since there is no self-consistent symmetries adopted here, the angular momentum projection demands to perform the integration over Euler angles $\Omega = (\alpha,\beta,\gamma)$ in full intervals without restrictions. To do so, we have developed an analytical formula that performs an exact integration over $(\alpha,\gamma)$ whose derivation is presented in Appendix B. The integration over $\beta$ is done numerically using the Gauss--Legendre quadrature rule.

\subsection{DNO-SM formalism}

\subsubsection{Secular equation in non-orthogonal PCHF basis}
Let us start with the ansatz~\eqref{eq:non_orthogonal_expansion} where we specify the eigenstate of $\hat{\mathcal H}$ by $\ket{\alpha JM}$ of good total angular momentum $J$, its projection $M$ in the laboratory frame and an index $\alpha$ labelling indices of corresponding energy levels and other quantum numbers. In terms of the PCHF states $\mathcal P^J_{MK}\ket{\Phi(q)}\in\mathit\Gamma_0\subset\Gamma$, it is given by
\begin{equation}
  \ket{\alpha JM} = \sum_{q,K} f^{(J)}_\alpha(q;K) \: \mathcal P^{J}_{MK}\ket{\Phi(q)}.
  \label{eq:alpha_JM_PCHF}
\end{equation}
To simplify the notation, we omit the dependence on mass number $A$ in the projection operator $\mathcal P^J_{MK}$ and $q$ is understood to take discrete values. The projected equation~\eqref{eq:projected_Hq_equation} then becomes
\begin{equation}
  \sum_{q,K}\bigg[\mathcal H^{J}_{K'K}(q',q) - E^{(J)}_\alpha\mathcal N^J_{K'K}(q',q)\bigg]
  f^{(J)}_\alpha(q;K) = 0
  \label{eq:projected_AMP_equation}
\end{equation}
where the Hamiltonian and the norm matrix elements $\mathcal O^{(J)}_{K'K}(q',q) = \elmx{\Phi(q')}{\hat{\mathcal O}\,\mathcal P^J_{K'K}}{\Phi(q)}$ (with $\hat{\mathcal O} = \hat{\mathcal H}, \mathbf 1$) are evaluated through a three-fold integration over Euler angles $\Omega = (\alpha,\beta,\gamma)$
\begin{equation}
  \begin{aligned}
    \mathcal O^{(J)}_{K'K}(q',q) &= \frac{2J+1}{4\pi^2\big(3-(-)^A\big)} \times \\
    &\int d\Omega \,\mathcal D^{J*}_{MK}(\Omega) \: \elmx{\Phi(q')}{\hat{\mathcal O}\hat R(\Omega)}{\Phi(q)}.
  \end{aligned}
  \label{eq:TBME_PCHF}
\end{equation}
The core of~\eqref{eq:TBME_PCHF} is in the evaluation of the kernels $\elmx{\Phi(q')}{\hat{\mathcal O}\hat R(\Omega)}{\Phi(q)}$ for given pair of Slater determinants. For these calculations, we use the minor formula as presented in Ref.~\cite{Watt1972}. This is given in Appendix A. The matrix element $\mathcal O^{(J)}_{K'K}(q',q)$ obtained from the Euler angles integration is then presented in Appendix B.

The treatment of the generalized eigenvalue problem~\eqref{eq:projected_AMP_equation} has been well documented in the framework of the generator coordinate method (see e.g. Refs.~\cite{GCMtriaxI,GCMtriaxII,GCMtriaxIII}). We follow the standard technique that begins with the determination of the natural eigenbasis functions $u^{(J)}_i(q;K)$ of the norm matrix
\begin{equation}
  \sum_{q,K}\mathcal N^J_{K'K}(q',q) \: u^{(J)}_i(q;K) = \eta^{(J)}_i \:u^{(J)}_i(q';K').
\end{equation}
By retaining only positive norm eigenvalues that we denote by $\{\eta^{(J)}_i > 0,\:i\in\mathbb N\}$, the so-called natural state characterized by the corresponding norm eigenvalue $\eta^{(J)}_i$ is defined as
\begin{equation}
  \begin{aligned}
    \ket{\eta^{(JM)}_i} = \frac{1}{\sqrt{\eta^{(J)}_i}} \sum_{q,K} u^{(J)}_i(q;K)
    \mathcal P^{J}_{MK}\ket{\Phi(q)},\\
    \overlap{\eta^{(JM)}_{i'}}{\eta^{(JM)}_i} = \delta_{i'i} \phantom{\mathcal P^{J}_{MK}\ket{\Phi(q)}}
  \end{aligned}
\end{equation}
and satisfies the orthogonality condition. This natural basis allows to transform the projected equation in the non-orthogonal PCHF basis onto the usual eigenvalue value problem of the form
\begin{equation}
  \sum_{i'}\mathcal H^{(J)}_{i'i} \: g^{(J)}_{i'} = E^{(J)}_\alpha \: g^{(J)}_i
\end{equation}
where the Hamiltonian matrix now is expressed between orthogonal natural basis states $\{\ket{\eta^{(JM)}_i}\}$
\begin{equation}
  \begin{aligned}
    \mathcal H^{(J)}_{i'i} = \frac{1}{\sqrt{\eta^{(J)}_{i'}\eta^{(J)}_i}}
    \sum_{q'K',qK}\: u^{(J)*}_{i'}(q';K') \times \\
    \mathcal H^J_{K'K}(q',q) \: u^{(J)}_i(q;K).
  \end{aligned}
\end{equation}

The nuclear state $\ket{\alpha JM}$ is thus a linear superposition in the natural basis
\begin{equation}
  \ket{\alpha JM} = \sum_{i}\; g^{(J)}_i \: \ket{\eta^{(JM)}_i}
\end{equation}
with the transformation onto the non-orthogonal PCHF basis expressed through the expansion coefficient $f^{(J)}_\alpha(q;K)$ of~\eqref{eq:alpha_JM_PCHF}
\begin{equation}
  f^{(J)}_\alpha(q;K) = \sum_{i}\; \frac{g^{(J)}_i}{\sqrt{\eta^{(J)}_i}} \: u^{(J)}_i(q;K).
\end{equation}

One can notice that $f^{(J)}_\alpha(q;K)$ is not properly normalized and does not represent the probability amplitude of finding a given configuration of $\mathit\Gamma_0$. To be able to analyze the content of the nuclear states, we can define the normalized probability amplitude to find a component $\mathcal P^J_{MK}\ket{\Phi(q)}$ in the state $\ket{\alpha JM}$ by~\cite{Ring1980,GCMtriaxI}
\begin{equation}
  \mathcal M^{(J)}_\alpha(q;K) = \sum_{q',K'}\; \big[\hat{\mathcal N}^{1/2}\big]^{(J)}_{K'K}(q',q)
  \: f^{(J)}_\alpha(q';K').
\end{equation}
The corresponding probability to find the intrinsic angular momentum component $K$ or the component $q$ in the state $\ket{\alpha JM}$ is respectively given by
\begin{equation}
  \begin{aligned}
    \mathcal P^{(J)}_\alpha(K) = \sum_{q} \Big|\mathcal M^{(J)}_\alpha(q;K)\Big|^2,\\
    \mathcal P^{(J)}_\alpha(q) = \sum_{K} \Big|\mathcal M^{(J)}_\alpha(q;K)\Big|^2,
  \end{aligned}
  \label{eq:prob}
\end{equation}
with the normalization relation
\begin{equation}
    \sum_{K} \mathcal P^{(J)}_\alpha(K) 
    = \sum_{q} \mathcal P^{(J)}_\alpha(q) = 1.
\end{equation}

\subsubsection{Truncation of the PCHF basis with the generalized minimization technique}\label{sec:truncation}
We come now to the truncation of the discrete family $\mathit\Gamma_0$ with the minimization technique. The idea about how to choose states of $\mathit\Gamma_0$ is quite straightforward. That is, applying to the present case including the angular momentum projection, based on the selection of discrete values of the coordinate(s) $q$ which minimizes the energy, i.e. we let the Hamiltonian itself to choose what is the best state from a variational viewpoint. Furthermore, it is implicitly understood that the ground state or an excited state has to be chosen beforehand. Then the following iterative procedure can be implemented:

\begin{enumerate}
\item[1)] Fix the state $E^{(J)}_\alpha$ to be minimized with $\alpha$ indexing energy levels;
\item[2)] Define a searching region of the coordinate(s) $q$;
\item[3)] Start from the first point which can be chosen as the HF minimum;
\item[4)] Solve the projected Shell Model circular equation~\eqref{eq:projected_AMP_equation}
  over the whole searching region of $q$ to find the second state and proceed the same way
  in next iterations until the convergence of $E^{(J)}_\alpha$.
\end{enumerate}
More precisely, the convergence criterion is defined by the absolute energy gain $\Delta E^{(J)}_{\alpha}(k) = E^{(J)}_\alpha(k) - E^{(J)}_\alpha(k-1)$ at the iteration number $k$. If $\Delta E^{(J)}_\alpha(k) > \epsilon > 0$ we keep the CHF state, otherwise it is not retained. The minimization will stop when no more states are found in the searching region.

In practice, we observe that fixing $(\alpha,J)$ will exclude in the minimization the states of $\mathit\Gamma_0$ which could be relevant for an other state $(\alpha',J')$. Hence, small components of the wave functions could be missed whereas the overall spectrum remains well described. The key point is thus, minimizing as many excited states at the same time as possible will eventually lead to the improvement of one and another mutually and also the ground state. This idea leads us to generalize the above iterative procedure in the following way: we let the minimization process to determine not only the coordinate(s) $q$ but also the state $(\alpha,J)$ that gains the most energy at a given iteration. This is done by comparing the energy differences $\{\Delta E^{(J)}_\alpha\}$ for every state $(\alpha,J)$ in a given set of nuclear states which we want to describe.
The procedure then continues until we find no more states satisfying the condition $\mathrm{max}\{\Delta E^{(J)}_\alpha\} > \epsilon$, which defines the convergence criterion in this scheme. We will comment on the choice of $\epsilon$ in the following.


The minimization procedure as such requires an organizational scheme of partitioning the coordinate(s) $q$, which form in general a multi-dimensional surface. Although it is possible to determine all of them simultaneously in principle, it might not be necessary to do so. The reason is that, as a consequence of the existence theorem, different countable sets $\mathit\Gamma_0\subset\Gamma$ could be qualified as basis for $\mathscr H_q$. In this work, we define the following organization of the coordinate(s) $q$. Only the deformation parameters $Q_{\lambda\mu}$ are determined from the minimization process. The cranking components $\langle\hat J_m\rangle$ ($m=x,z$) are not. Instead, they are fixed in advance. That is, we perform the minimization in the deformation surface $\{Q_{\lambda\mu}\}$ associated to each value of $\langle\hat J_m\rangle = J^{(1)}_m,J^{(2)}_m,J^{(3)}_m,...$ ordered from input.


\subsubsection{Choice of the coordinate(s) $q$}\label{Choice of the coordinate(s) $q$}

In order the generate $\mathit\Gamma_0$, in this present work, we limit ourselves to quadrupole deformations (axial and triaxial) as the common choice of generator coordinates for triaxial systems~\cite{Ring1980}, whose expectation values in a HF state $\ket\Phi$ are denoted by $Q_{20},Q_{22}$ respectively
\begin{align}
  Q_{20} &= \sqrt{\frac{16\pi}{5}}\, \sum_{\tau=p,n}\:e^{(\tau)}_{mass}\:\elmx{\Phi}{\hat Q^{(\tau)}_{20}}{\Phi} \\
  Q_{22} &= \sqrt{\frac{8\pi}{5}}\, \sum_{\tau=p,n}\:e^{(\tau)}_{mass}\:
  \elmx{\Phi}{\big(\hat Q^{(\tau)}_{2-2} + \hat Q^{(\tau)}_{22}\big)}{\Phi}.
\end{align}
In this particular case, we use the usual Hill-Wheeler $(\beta,\gamma)$ parameters~\cite{betagamma} which are related to $(Q_{20},Q_{22})$ through the total quadrupole moment $Q=\sqrt{Q_{20}^2 + Q_{22}^2}$
\begin{equation}
  \beta = \frac{b^2Q\sqrt{5\pi}}{3r_0^2A^{5/3}}, \:\:
  \gamma = \arctan\big(\frac{Q_{22}}{Q_{20}}\big)
\end{equation}
where $b^2$ (in fm$^2$) is the harmonics oscillator parameter~\cite{Brussard1977,js07,NilssonSU3}
\begin{equation}
  b^2 = \frac{41.4}{45A^{-1/3} - 25A^{-2/3}},
\end{equation}
and $A$ and $r_0=1.2$ (in fm) are the nuclear mass number and radius parameter.

Within the shell-model formalism, the use of valence spaces and truncation of the Hilbert space implies
the need of effective hamiltonians as well as effective operators. Effective operators can be derived 
by Many-Body Perturbation Theory~\cite{Coraggio:2020cmw} but quadrupole operators are usually renormalized with the use of an effective charge. The effective charge have been defined by several authors, we use here the notation defined in Ref.~\cite{js07}: 
\begin{equation}
e_{el}^{(p)}=(1+\chi_p)e, \:\:\: e_{el}^{(n)}=\chi_n e
\end{equation}
with $\chi$ being the electric polarization charge. 
$\chi$ value is intimately connected to the valence space used and can be derived by Many-Body Perturbation Theory~\cite{Coraggio:2020cmw} and is shell dependent but for $0\hbar\omega$ spaces, the microscopic Dufour-Zuker~\cite{hmulti} $\chi_p=0.31$, $\chi_n=0.46$ are standard values. Finally, for the mass quadrupole operators, we will use 
\begin{equation}
e_{mass}^{(p)} = e_{mass}^{(n)} = (1+\chi_p+\chi_n)e
\label{eq:mass_charge}
\end{equation}
to be consistent with the deformation parameters defined above.


\section{Benchmark and comparison in sd nuclei} \label{Benchmark and comparison in sd nuclei}
\subsection{Even nuclei}

To assess the quality of the PCHF basis $\mathit\Gamma_0$  against the oscillator one in the classic Shell Model, we perform a systematic comparison of two models in even sd nuclei using the USDB effective interaction~\cite{Brown:USDB}. 

\begin{figure}[H]
  \includegraphics[scale=0.45]{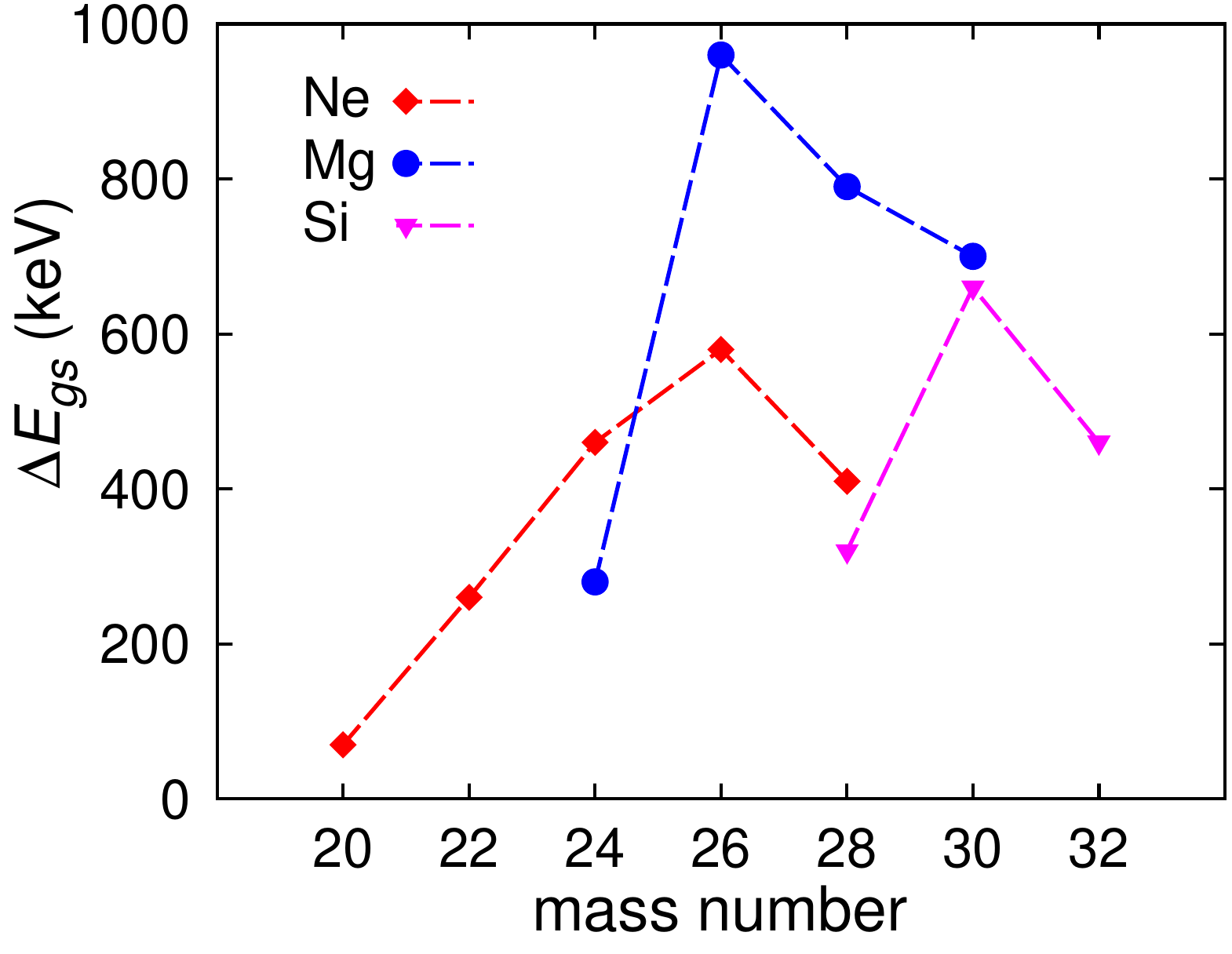}
  \caption{\label{Egs} Energy difference (in keV) between the DNO-SM and the exact SM diagonalizations in the ground state of even Ne, Mg and Si isotopes from Table~\ref{TAB_SD}.}
\end{figure}

In the first series of calculations shown in Table~\ref{TAB_SD}, to see the overall performance of our minimization method, we compare the ground state and the first excited states $2^+_1,4^+_1$ energies as well as the corresponding $E2$--transition probabilities for Neon, Magnesium, Silicon and Argon even-even isotopes. These calculations are performed using: i) the common set of states $J_\alpha\in\{0_1,2_{1,2},4_1,8_{1},12_1,16_1\}$ whenever possible; ii) practically the same $(\beta,\gamma)$ mesh for all considered nuclei (cf. Table~\ref{TAB_SD_BG}); iii) the same set of cranking components $\langle\hat J_z\rangle\in\{-2,-4,-8,-12,-16\}$ and iv) the same number of integration points in Angular Momentum Projection, namely, $11$ points for the analytical integration over the angles $\alpha,\gamma\in[0,2\pi]$ and $20$ Gauss--Legendre quadrature points for the integration over the angle $\beta\in[0,\pi]$. After a first round of minimization to find non-cranked CHF states over the range $\gamma\in[0^\circ,60^\circ]$ discretized into a one-dimensional mesh of $N_g$ points, the minimization procedure is then iterated for each above-mentioned cranking components $\langle\hat J_z\rangle$.\\
As we can see in Table~\ref{TAB_SD}, the agreement between two models is excellent for both relative energies and $E2$--transition probabilities, indicating that the wave functions have already converged for the considered states. The number of CHF basis states found by the minimization process is relatively small at the beginning and the end of the shell and increases as we go to nuclei at mid shell. Such are the cases of deformed open shell nuclei $^{24}$Mg, $^{28}$Si which seemingly require more basis states to capture correlations in the mixing than in others. In the present calculations, a similar trend is exhibited in the ground state binding energy difference (of order~$\sim 0.1-1.0$ MeV) with respect to the exact result (shown in Figure~\ref{Egs}). There are two possible reasons for this underbinding:\\
$\bullet$ Our minimization technique builds up the many-body basis based on a finite set of states preliminarily defined, in the present case, i.e. the set $J_\alpha\in\{0_1,2_{1,2},4_1,8_{1},12_1,16_1\}$, hence, indirectly omits basis states possibly relevant for other states not included in the set. Therefore, the representation of the effective Hamiltonian in the truncated basis $\mathit\Gamma_0$ is not fully complete as the construction is under way. This is in contrast to the full SM diagonalization in the oscillator basis because all many-body matrix elements of the effective Hamiltonian are available once the valence space is defined.\\
$\bullet$ The second reason may be related to the choice of the coordinate(s) $q$. Here, the question that arises is which $q$ to be used to ensure \textit{a priori} the full representation of the effective Hamiltonian.

\begin{table}[H]
  \centering
  \caption{\label{TAB_SD_BG} Discretization of the $(\beta,\gamma)$ plane for minimization procedure for selected sd nuclei using the USDB effective interaction. The Hartree-Fock minimum energy $(E_{\rm HF})$ and corresponding mass quadrupole parameters $(\beta_{\rm min},\gamma_{\rm min})$ are also given. $\gamma_{\rm min}$ is in degree. The effective charges for protons and neutrons are $e^{(p)}_{mass}=e^{(n)}_{mass}=1.77$ (see EQ.\eqref{eq:mass_charge}).}
  \begin{tabular}{*{7}c}
    \hline\hline
    $\beta$ & $N_b$ & $N_g$ & nucleus   & $\beta_{\rm min}$ & $\gamma_{\rm min}$ ($^{\circ}$) & $E_{\rm HF}$ (MeV) \\
    \hline
    \multirow{2}{*}{$[0.1,0.7]$} & \multirow{2}{*}{$11$} & \multirow{2}{*}{$9$} &  $^{20}$Ne & $0.527$ & $0.0$ & $-36.404$\\
            &       &      &  $^{22}$Ne & $0.498$ & $0.0$ & $-53.474$\\ \hline 
     $[0.1,0.53]$ & $11$ & $9$ &  $^{24}$Ne & $0.313$ & $23.8$ & $-66.225$\\ \hline
    \multirow{10}{*}{$[0.1,0.48]$} & \multirow{10}{*}{$7$} & \multirow{10}{*}{$9$} &  $^{26}$Ne & $0.247$ & $0.0$ & $-77.258$\\
            &       &      &  $^{28}$Ne & $0.198$ & $0.0$ & $-83.441$\\
            &       &      &  $^{24}$Mg & $0.499$ & $12.0$ & $-80.965$\\
            &       &      &  $^{26}$Mg & $0.395$ & $34.0$ & $-98.887$\\
            &       &      &  $^{28}$Mg & $0.325$ & $0.0$ & $-115.625$\\
            &       &      &  $^{30}$Mg & $0.212$ & $0.0$ & $-126.735$\\
            &       &      &  $^{28}$Si & $0.425$ & $60.0$ & $-130.021$\\
            &       &      &  $^{30}$Si & $0.285$ & $46.8$ & $-148.238$\\
            &       &      &  $^{32}$Si & $0.221$ & $60.0$ & $-166.344$\\
            &       &      &  $^{32}$S  & $0.0$  &  $-$ & $-176.393$\\
            &       &      &  $^{34}$S  & $0.152$ & $60.0$ & $-198.097$\\
            &       &      &  $^{36}$Ar  & $0.199$ & $60.0$ & $-226.611$\\
    \hline\hline
  \end{tabular}
\end{table}
\begin{table*}[t]
  \caption{Systematics comparison of the diagonalization in the model spaces $\mathscr H_q$ and $\mathscr H$ of the Shell Model in even sd nuclei. The absolute ground state energy $E_{\rm gs}$ and relative energies of the first $2^+$ and $4^+$ are given in MeV. Their reduced transition probabilities $B(E2;2^{+}_1\to 0^+_1)$ and $B(E2;4^{+}_1\to 2^+_1)$ are in e$^2$.fm$^4$ unit. $N_q$ denotes the number of CHF states $\ket{\Phi(q)}$ found with the minimization procedure.\label{TAB_SD}}
\begin{ruledtabular}  
  \begin{tabular}{*{17}c}
    \multirow{2}{*}{nucleus} && \multicolumn{3}{c}{$\mathscr H_q$} &&
    \multicolumn{2}{c}{$\mathscr H$} && \multicolumn{2}{c}{$E_{\rm gs}$ (MeV)} &&
    \multicolumn{2}{c}{$B(E2;2^{+}_1\to 0^+_1)$} &&
    \multicolumn{2}{c}{$B(E2;4^{+}_1\to 2^+_1)$}  \\
    \cline{3-5} \cline{7-8} \cline{10-11} \cline{13-14}\cline{16-17}
    &&  $N_q$ & $2^+_1$ & $4^+_1$  &&
    $2^+_1$ & $4^+_1$  && $\mathscr H_q$ & $\mathscr H$ &&
    $\mathscr H_q$ & $\mathscr H$ && $\mathscr H_q$ & $\mathscr H$   \\
    \hline
    $^{20}$Ne && $16$ & $1.76$ & $4.14$ && $1.75$ & $4.18$ && $-40.40$ & $-40.47$ && $47.0$ & $46.9$ && $56.3$ & $55.3$   \\
    $^{22}$Ne && $41$ & $1.37$ & $3.39$ && $1.36$ & $3.36$ && $-57.32$ & $-57.58$ && $48.1$ & $46.9$ && $64.0$ & $63.3$   \\
    $^{24}$Ne && $39$ & $2.13$ & $4.01$ && $2.11$ & $3.99$ && $-71.26$ & $-71.72$ && $38.8$ & $38.7$ && $31.9$ & $31.3$   \\
    $^{26}$Ne && $26$ & $2.12$ & $3.72$ && $2.06$ & $3.51$ && $-80.98$ & $-81.56$ && $39.6$ & $38.5$ && $37.8$ & $33.8$   \\
    $^{28}$Ne && $12$ & $1.55$ & $2.82$ && $1.62$ & $2.99$ && $-86.13$ & $-86.54$ && $34.3$ & $34.1$ && $32.7$ & $30.7$ \\
    $^{24}$Mg && $42$ & $1.52$ & $4.37$ && $1.50$ & $4.37$ && $-86.82$ & $-87.10$ && $76.1$ & $74.4$ && $99.1$ & $97.5$\\
    $^{26}$Mg && $50$ & $1.80$ & $4.36$ && $1.89$ & $4.36$ && $-104.56$ & $-105.52$ && $66.1$ & $65.2$ && $27.5$ & $18.0$ \\
    $^{28}$Mg && $50$ & $1.40$ & $4.07$ && $1.52$ & $4.17$ && $-119.71$ & $-120.50$ && $62.9$ & $60.2$ && $75.7$ & $67.5$ \\
    $^{30}$Mg && $21$ & $1.50$ & $3.89$ && $1.59$ & $3.89$ && $-129.77$ & $-130.47$ && $53.0$ & $49.1$ && $39.6$ & $32.5$  \\
    $^{28}$Si && $71$ & $2.12$ & $4.78$ && $1.93$ & $4.61$ && $-135.54$ & $-135.86$ && $77.4$ & $77.9$ && $106.8$ & $109.6$ \\
    $^{30}$Si && $46$ & $2.25$ & $5.59$ && $2.26$ & $5.33$ && $-154.09$ & $-154.75$ && $46.9$ & $45.9$ && $54.0$ & $15.8$ \\
    $^{32}$Si && $43$ & $1.99$ & $5.79$ && $2.05$ & $5.88$ && $-170.06$ & $-170.52$ && $41.1$ & $42.5$ && $66.9$ & $65.2$ \\
    $^{32}$S  && $50$ & $2.05$ & $4.55$ && $2.16$ & $4.65$ && $-182.10$ & $-182.45$ && $48.0$ & $46.9$ && $69.2$ & $66.8$ \\
    $^{34}$S  && $36$ & $2.01$ & $4.66$ && $2.13$ & $4.83$ && $-202.08$ & $-202.50$ && $38.3$ & $36.0$ && $51.7$ & $48.0$ \\
    $^{36}$Ar && $12$ & $1.81$ & $4.46$ && $1.82$ & $4.49$ && $-230.22$ & $-230.28$ && $50.4$ & $50.6$ && $63.2$ & $62.9$ \\
  \end{tabular}
\end{ruledtabular}  
\end{table*}
\subsection{Detailed analysis of $^{24}$Mg}
In the second series of calculations, we investigate the specific case of $^{24}$Mg, which is a typical example of deformed nuclei in the sd shell. This nucleus was extensively studied in the past within various approaches~\cite{GCMtriaxI,GCMtriaxII,Mg24Yao,zch09,Mg24YaoII}. It is well established that it has a triaxial shape in its ground state with a rotational band built on top. Moreover, the triaxiality manifests itself in the existence of the so-called $\gamma$--band that was experimentally observed. In order to examine whether our method is able to correctly describe these features, we explore different minimization schemes as proposed in~\ref{sec:truncation}.

To fix ideas, let us first mention that the present calculations of $^{24}$Mg as shown in Figure~\ref{Mg24spec} are performed using the same $(\beta,\gamma)$ discretization as in Table~\ref{TAB_SD_BG} and the cranking components $\langle\hat J_z\rangle \in\{0,-2,-3,-4,-5\}$. Results reported in the figures~\ref{Mg24spec}b),~\ref{Mg24spec}c) and~\ref{Mg24spec}d) are obtained from the ground-state and many states minimizations in comparison with the full mesh diagonalization~\ref{Mg24spec}e) and the exact SM diagonalization~\ref{Mg24spec}f). The searching domain for minimizations is also the dimension of the full mesh calculation which is \textit{a priori} $7\times 9\times 5 = 315$. In practice, our Hartree-Fock calculations found $277$ converged solutions.

From Figure~\ref{Mg24spec} the first thing we note is that Angular Momentum Projection from the HF minimum gives already very good intraband transitions. In contrast, the interband transitions $BE(2;2^+_2\rightarrow 2^+_1)$ or $BE(2;6^+_1\rightarrow 5^+_1)$ which reflect the $K$-mixing of the two bands are significantly underestimated. In minimizing the ground state alone, the interband transitions are considerably improved and consistent with the exact SM result using only $6$ basis states. This means we have already the right wave functions with the mixing of a few number of CHF states, even though about $\sim 600$ keV is still missing in the ground state binding. Carrying out the same minimization of the ground state until no more CHF basis vectors are found in the full mesh, we obtain $39$ states that perfectly reproduce the exact SM result. This calculation also confirms that in the ground-state minimization, $6$ CHF vectors is enough to have good solutions in this nucleus.

Now to see whether we can improve further by minimizing many states at the same time, as shown in the figure~\ref{Mg24spec}d), we find $55$ CHF states that provide an overall better energy spectra than the ground-state minimization but does not change much the picture. The excited states in the figure~\ref{Mg24spec}d) are slightly more compressed. This effect, as shown in Figure~\ref{cranking_comparison}, can be traced back to the addition of cranked CHF states which are known to stretch down the relative spectrum, although in the present calculation the effect is not as huge as noted in e.g. Ref.~\cite{Tomas2015cranking}.

\begin{figure}[H] 
  \includegraphics[scale=0.22]{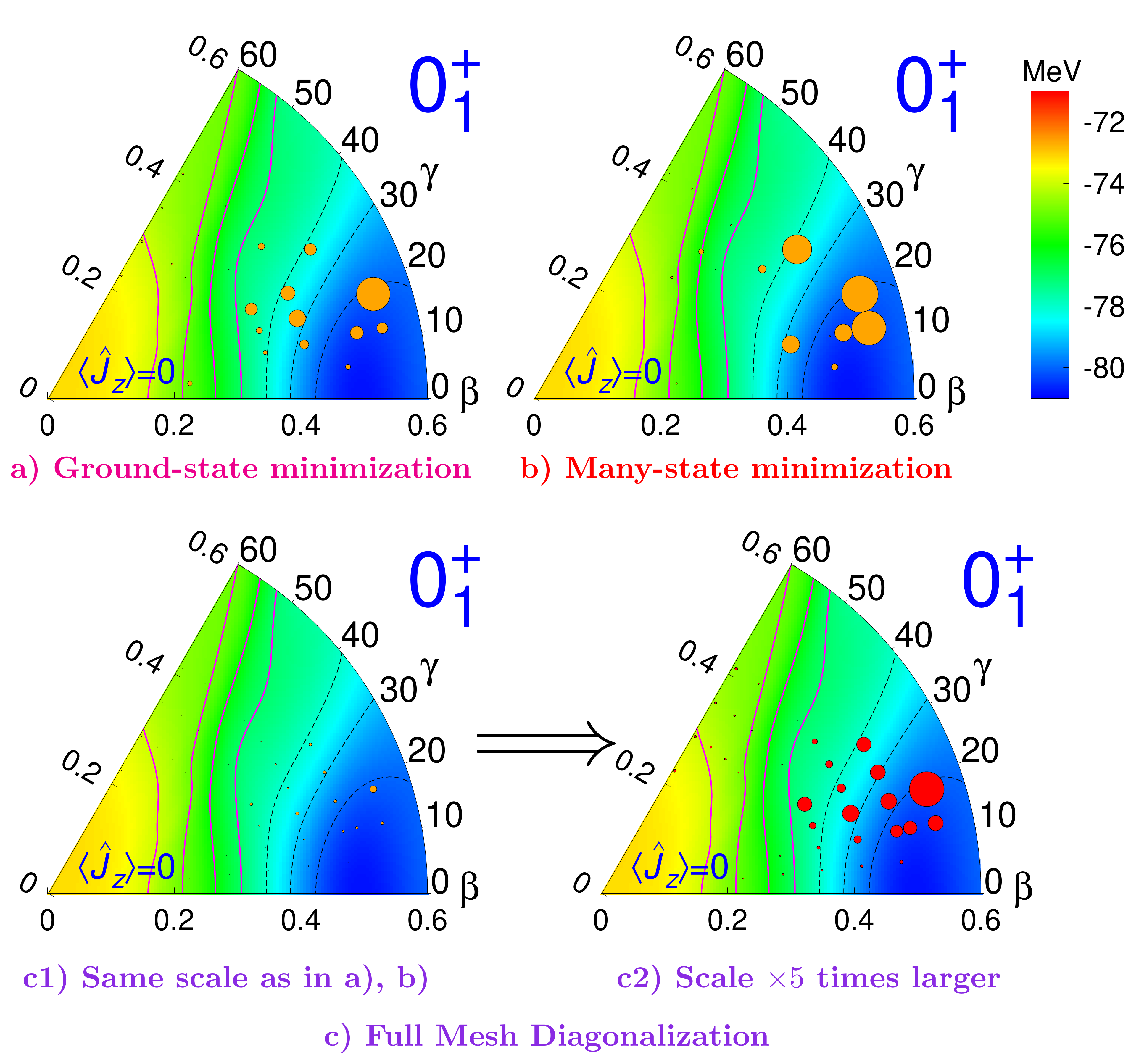}
  \caption{\label{Mg24CHF0+1} Structure of the ground state depicted in the potential energy surface of $\langle\hat J_z\rangle = 0$. The yellow circles (defined with the same scale) in a), b) and c1) represent the contribution of CHF basis vectors used in the ground-state minimization, the many-state minimization and the full mesh diagonalization respectively. In c2) the red circles represent the same CHF basis vectors of c1) but using a scaling factor of $5$ times larger (see discussion in texts).}
\end{figure}

\begin{center}
\begin{figure}[H]
\includegraphics[scale=0.27]{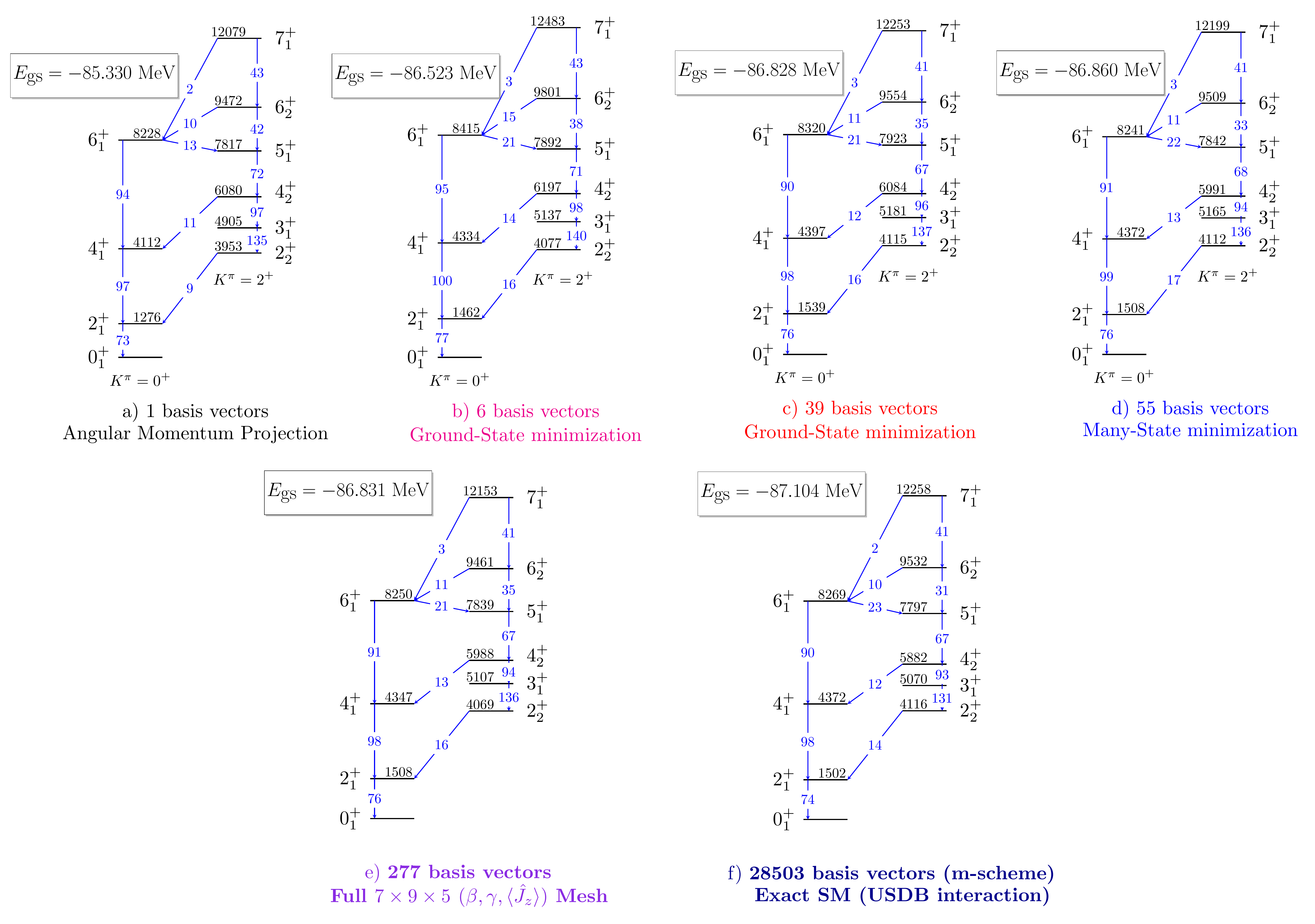} 
\caption{\label{Mg24spec} $^{24}$Mg spectrum calculated with the USDB interaction using the DNO-SM compared to the classic SM diagonalization. Black numbers are relative energies (in keV) and blue ones the $B(E2)$ values (in e$^2$.fm$^4$). $K$ denotes the dominant wave--function component of different members of the band. Figures on top show the evolution of the energy spectra between a) the Angular Momentum Projection of the Hartree-Fock minimum, b) and c) the minimization of the ground state, and d) the minimization of all considered states. Figures at bottom present e) the DNO-SM diagonalization in the full $7\times 9\times 5$ $(\beta,\gamma,\langle\hat J_z\rangle)$ mesh (see texts) and f) the exact SM diagonalization. The ground state binding energy $(E_{\rm gs})$ is also given for comparison.}
\end{figure}
\end{center}

The first difference between the two minimization schemes are in the absolute binding energies as reported in Table~\ref{TAB_Full_GS}.  Minimizing all the states provides solutions with lower bindings which are, in addition, in excellent agreement with the full mesh diagonalization. The second difference that we observe is in the state decomposition in the $(\beta,\gamma)$ plane. In Figures~\ref{Mg24CHF0+1}a),~\ref{Mg24CHF0+1}b) and~\ref{Mg24CHF0+1}c), for illustration, we show the contributions of CHF basis vectors into the ground state wave function in the potential energy surface of $\langle\hat J_z\rangle=0$. By comparing the figures~\ref{Mg24CHF0+1}a) and~\ref{Mg24CHF0+1}b), it is worth pointed out that the two minimizations produce two sets of basis vectors which give energy spectra and transitions of the same quality. The contributions of CHF states in the figure~\ref{Mg24CHF0+1}b) are somehow fragmented into other configurations in the figure~\ref{Mg24CHF0+1}a). The effect is even more visible in the full mesh calculation shown in the figure~\ref{Mg24CHF0+1}c1), which means there is a redistribution of contributions in the final wave function when one uses "redundant" basis states. Now, what we observe in the figure~\ref{Mg24CHF0+1}c2) is the key point: the dominant CHF configurations in the full mesh diagonalization are lying in the same region as in the minimization. In Table~\ref{TAB_CHF}, we report some of the most dominant configurations in the ground state as shown in Figure~\ref{Mg24CHF0+1}. The most important configurations are around the Hartree-Fock minimum and they appear in both minimization calculations. In the full mesh calculation, apart from a scaling effect due to the presence of many other states, it is the same CHF configurations which contribute the most. This clearly shows that our minimization method picks up the "right" physical configurations of the most importance. This comparison also explains the reason why the minimization using $6$ states gives already a good description.

We summarize now the essential points which can be drawn from the present discussion:\\
i) There can be different sets of CHF states that provide the same description of the relative energy spectra and transitions.\\
ii) There exists specific CHF states that contribute more significantly than others. And the minimization tells us where to find them.\\
iii) One must pay attention to the interpretation because of i), it is possible to replace some set of basis states with an other, although it could be at the price of taking into account more basis states than necessary.\\
iv) It is more advantageous to perform the minimization than the full mesh diagonalization since basis states are interdependent in the sense of iii). In cases where such calculation is feasible like in the present one, the full mesh diagonalization should expose the limit one obtains in the minimization. By the way, it is worth noting here that in a calculation from Monte-Carlo Shell Model of Ref.~\cite{MCSM1996Mg24} employing a stochastic sampling method to choose basis states, $800$ Hartree-Fock states stochastically chosen give $-86.91$ MeV of ground state binding for $^{24}$Mg (versus the exact value $-87.08$ MeV) with the USD interaction.\\
v) Finally, let us comment on a technical point regarding the choice of $\epsilon$ defining the convergence criterion of the method. For the present study, we use $\epsilon = 1$ keV. Such value seems to be too strict but is necessary to push the method to its limit and to see whether it reproduces the exact SM result. For that goal, this value is sufficient to avoid linear redundancies in the selection of CHF states. The redundancies among non-orthogonal states are very well known in the generator coordinate method and resulting from the nearly zero norm eigenvalues of the overlap matrix. In the minimization process, they manifest themselves through very tiny contributions in energy of order $\lesssim 10^{-1}$ keV. We therefore exclude basis states yielding those negligible contributions.

\begin{table}[H]
\begin{center}
\scalebox{1}{\begin{tabular}{*{6}c}
\hline\hline
\multirow{2}{*}{$\beta$} & \multirow{2}{*}{$\gamma$} && \multicolumn{3}{c}{$\mathcal P^{(J)}_\alpha(\beta,\gamma)$ ($\%$)} \\\cline{4-6}
&   &&  a) ground-state & b) Many-state & c) Full \\
\hline
$0.499$ & $12.0$ && $5.62$ & $7.43$ & $1.16$\\
$0.540$ & $17.74$ && $14.35$ & $15.58$ & $2.97$\\
$0.540$ & $11.85$ && $4.71$ & $14.64$ & $1.30$\\
$0.414$ & $17.74$ && $7.28$ & -       & $1.44$\\
$0.351$ & $23.63$ && $5.25$ & -       & $1.23$\\
$0.414$ & $11.85$ && $3.88$ & $7.54$  & $0.66$\\
$0.414$ & $29.52$ &&   -    & $3.39$  & $0.61$\\
$0.477$ & $17.74$ &&   -    & -    & $1.38$ \\
$0.477$ & $23.63$ &&   -    & -    & $1.29$ \\\hline\hline
\end{tabular}}
\end{center}
\caption{\label{TAB_CHF} Contributions of CHF basis states (calculated from~\eqref{eq:prob}) into the ground state wave function in three calculations shown in Figure~\ref{Mg24CHF0+1}a),~\ref{Mg24CHF0+1}b) and~\ref{Mg24CHF0+1}c) respectively.}
\end{table}

\begin{table}[H]
\begin{center}
\scalebox{0.9}{\begin{tabular}{*{7}c}
\hline\hline
&$J^\pi_\alpha$ & $0_1^+$ & $2_1^+$ & $4_1^+$ & $6_1^+$ & $8_1^+$ \\
&Figure~\ref{Mg24spec}c)   & $-86.828$ & $-85.288$ & $-82.430$ & $-78.508$ & $-75.028$ \\
&Figure~\ref{Mg24spec}d) & $-86.861$ & $-85.353$ & $-82.489$ & $-78.619$ & $-75.283$ \\
&Figure~\ref{Mg24spec}e) & $-86.831$ & $-85.322$ & $-82.483$ & $-78.580$ & $-75.249$ \\
\hline
\end{tabular}}
\scalebox{0.9}{\begin{tabular}{*{7}c}
$J^\pi_\alpha$ & $2_2^+$ & $3_1^+$ & $4_2^+$ & $5_1^+$ & $6_2^+$  & $7_1^+$\\
Figure~\ref{Mg24spec}c) & $-82.713$ & $-81.647$ & $-80.744$ & $-78.905$ & $-77.274$    & $-74.576$ \\
Figure~\ref{Mg24spec}d) & $-82.748$ & $-81.695$ & $-80.869$ & $-79.018$ & $-77.351$  & $-74.661$ \\
Figure~\ref{Mg24spec}e) & $-82.761$ & $-81.724$ & $-80.843$ & $-78.992$ & $-77.370$  & $-74.677$ \\
\hline
\end{tabular}}
\end{center}
\caption{\label{TAB_Full_GS} Comparison of binding energies (in MeV) obtained from the minimization of the ground state, of many states and the full mesh diagonalization in Figures~\ref{Mg24spec}c),~\ref{Mg24spec}d) and~\ref{Mg24spec}e) respectively.}
\end{table}

\onecolumngrid

\begin{center}
\begin{figure}[H]
  \scalebox{0.26}{\includegraphics{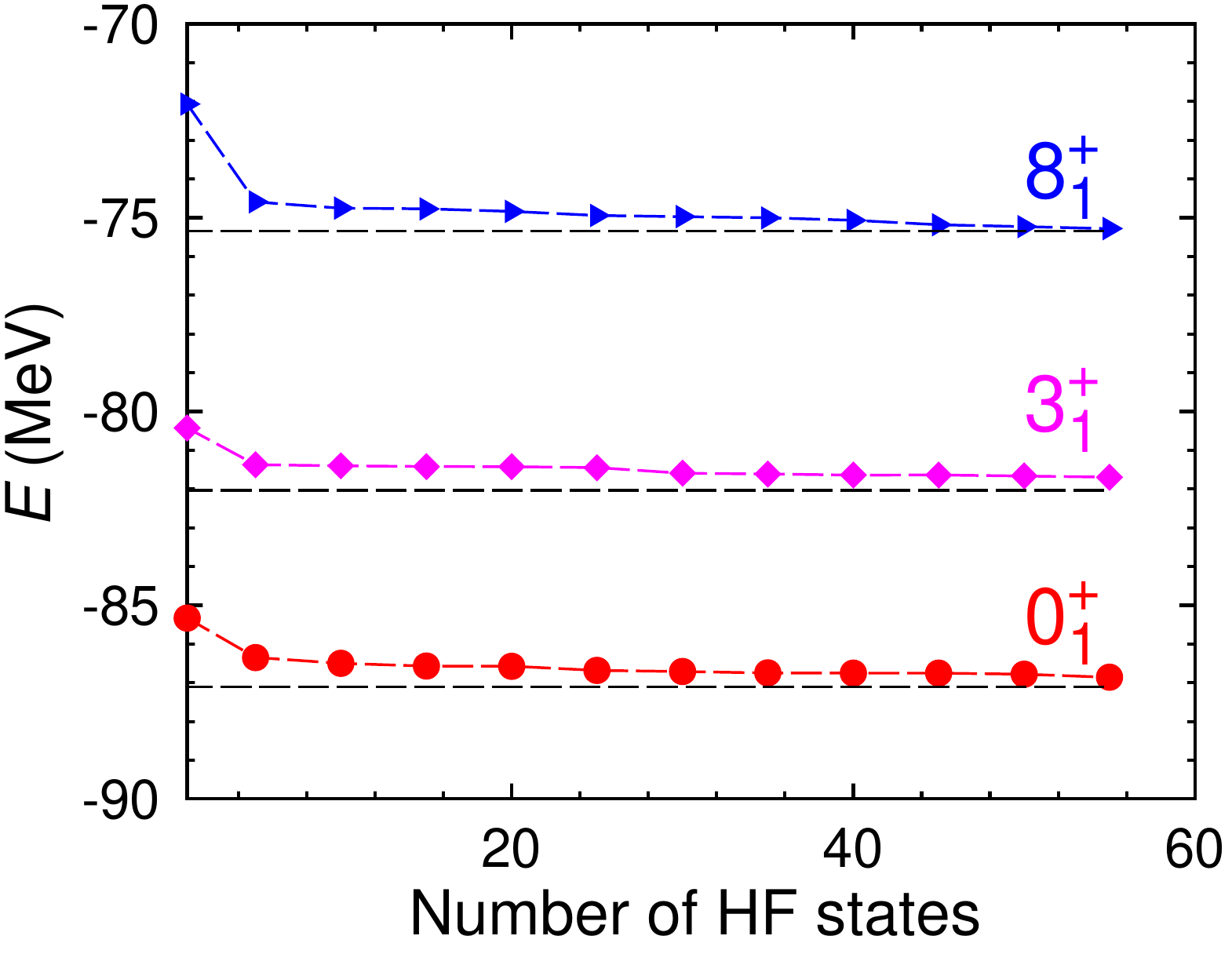}}
  \scalebox{0.26}{\includegraphics{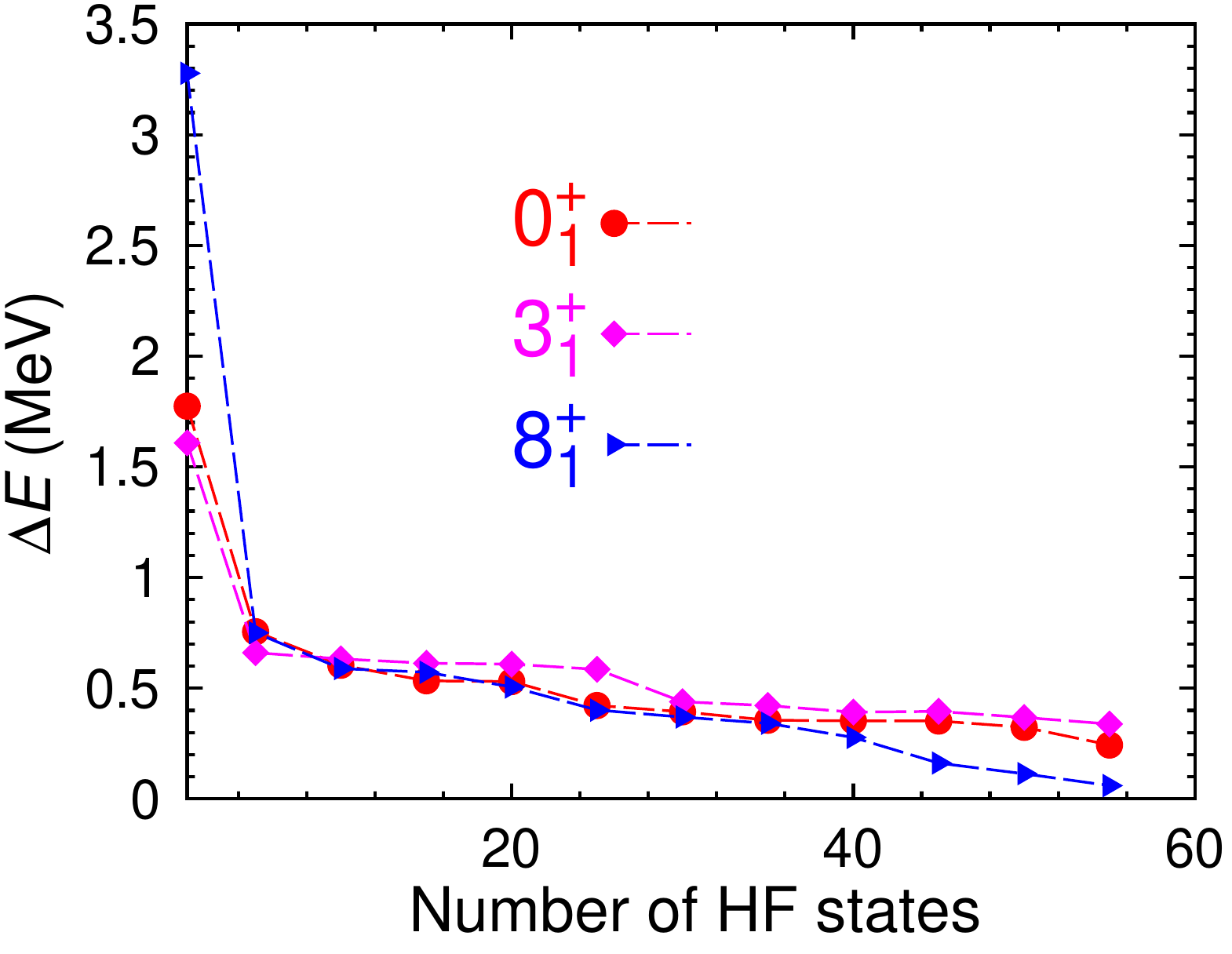}}
  \centering\scalebox{0.26}{\includegraphics{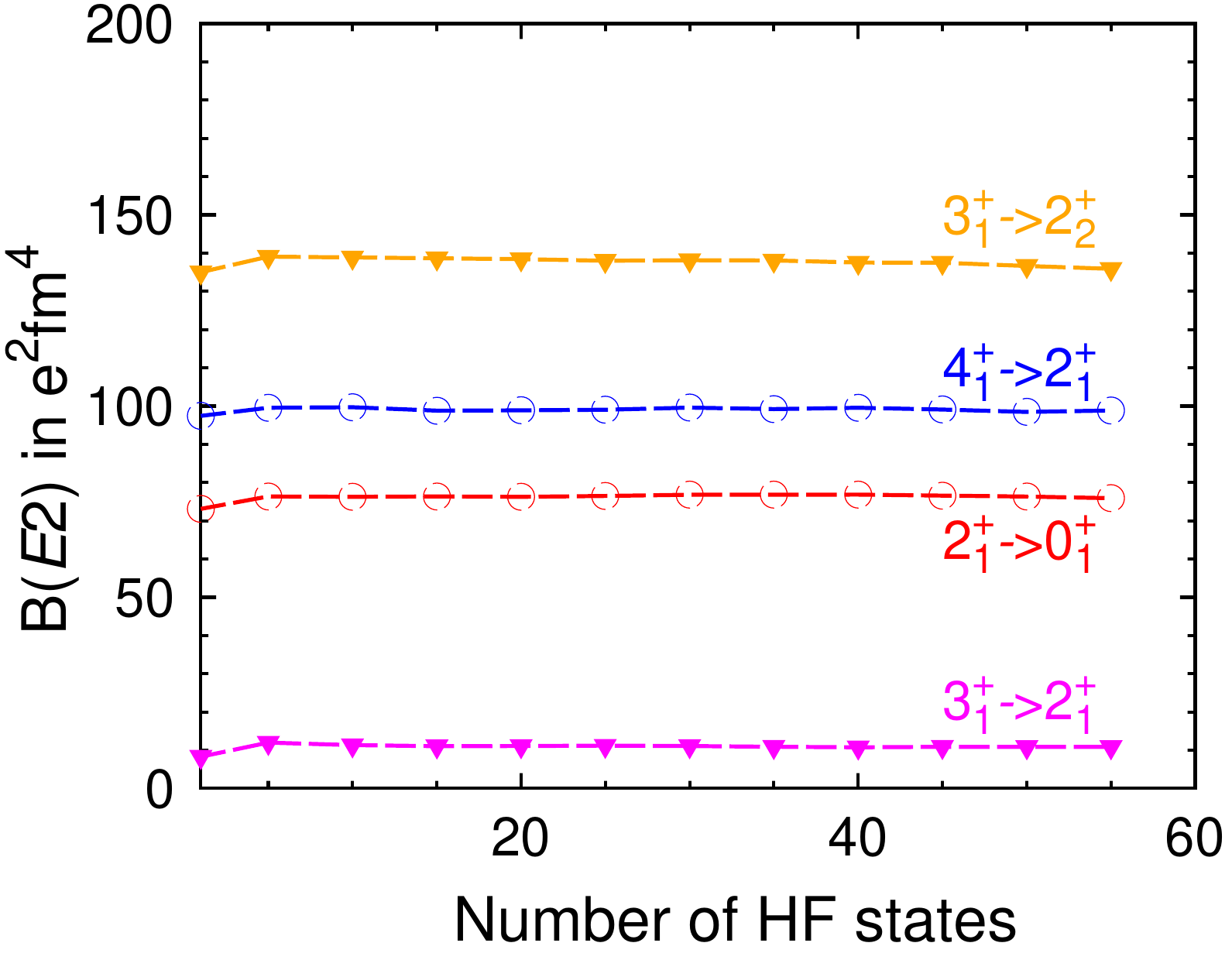}}
  \caption{\label{convergence} Convergence of the binding energies ($E$) (top left) where the horizontal lines represent the exact SM values, the binding energy difference ($\Delta E$) (top right) with respect to the exact SM result and the $B(E2)$ transitions (bottom) as a function of CHF states in the many-state minimization shown in Figure~\ref{Mg24spec}d).}
\end{figure}
\end{center}

\twocolumngrid

\begin{figure}[H]
  \scalebox{0.25}{\includegraphics{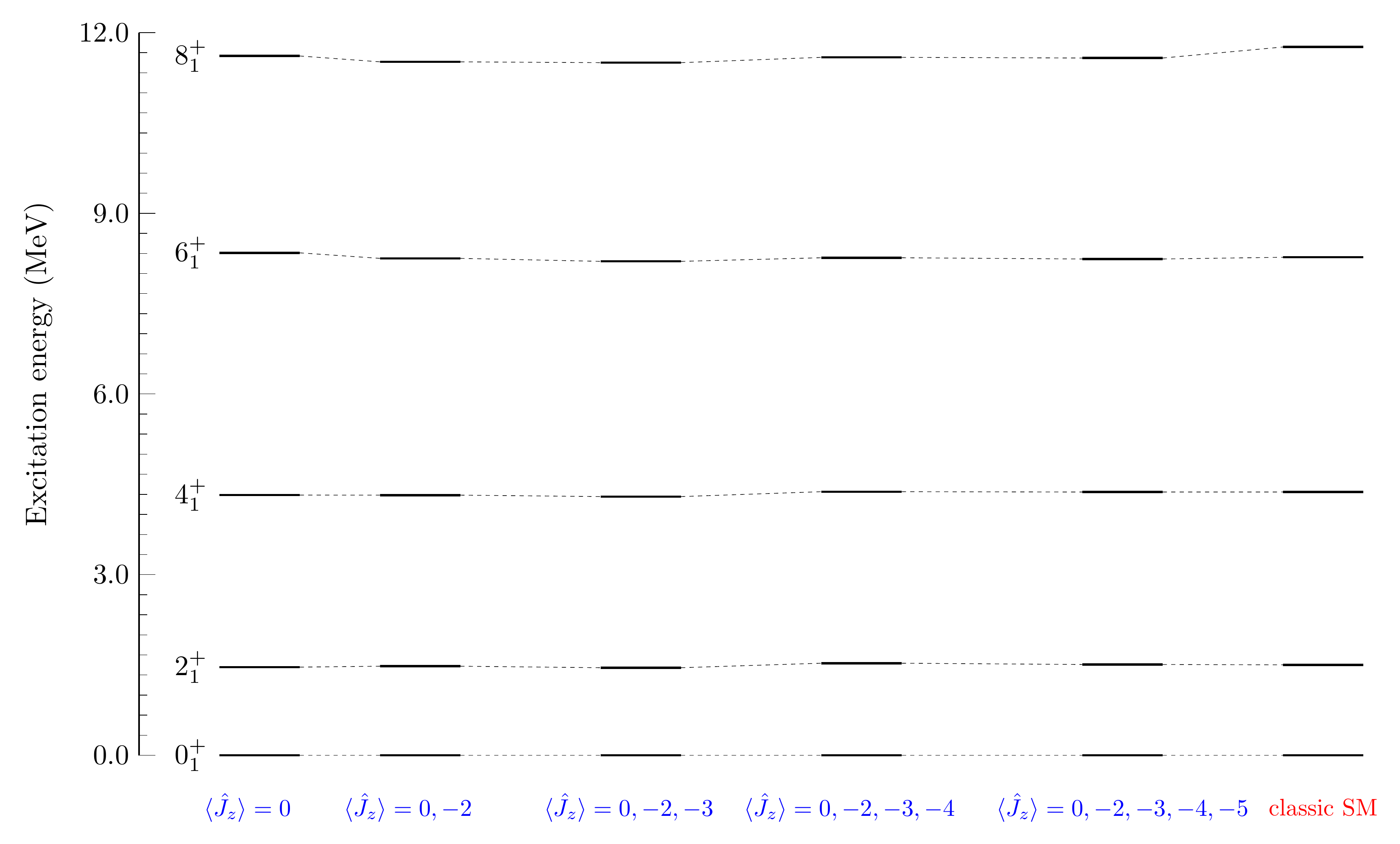}}
  \scalebox{0.25}{\includegraphics{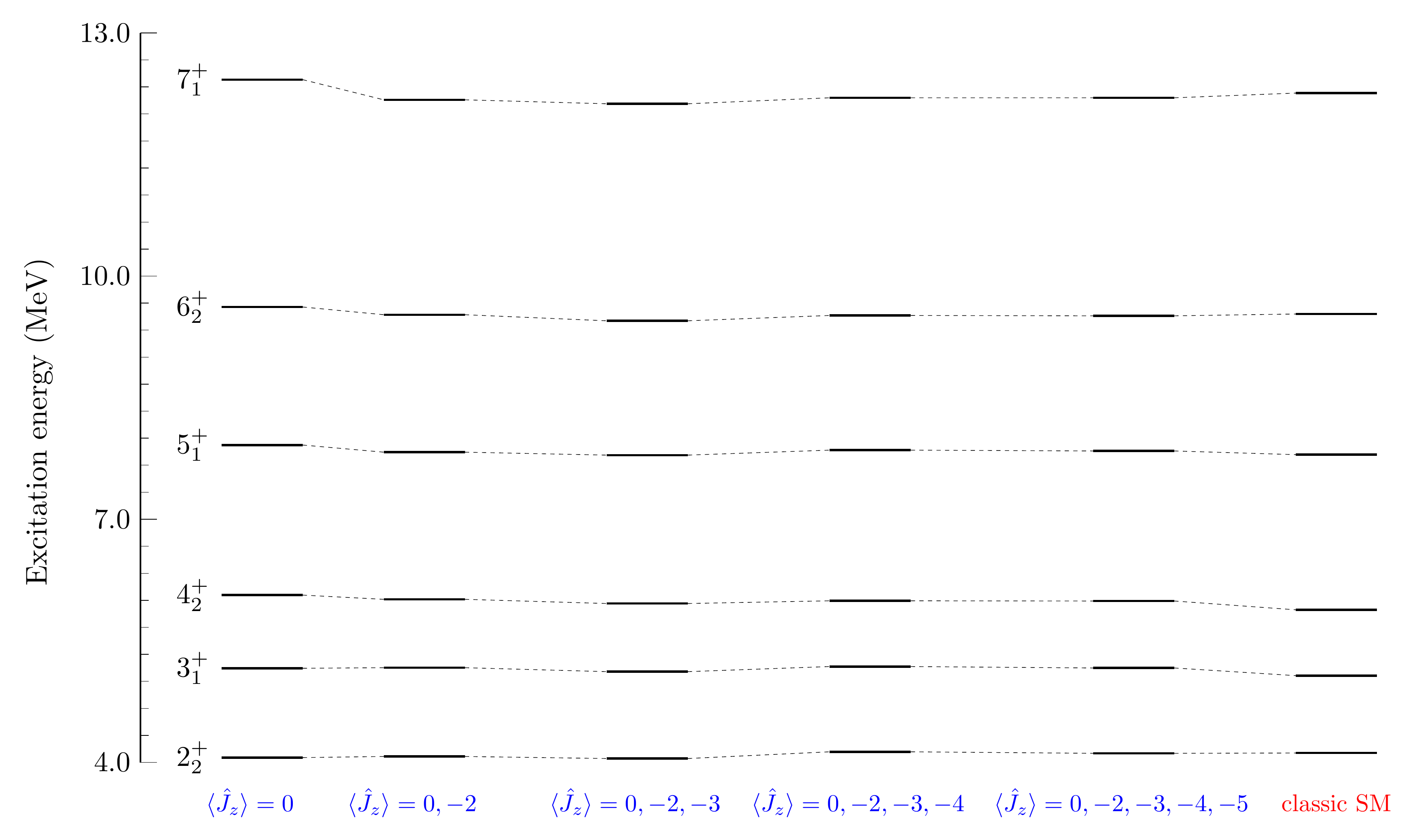}}
  \caption{\label{cranking_comparison} Evolution of the ground state band (top) and the $\gamma$-band (bottom) in $^{24}$Mg as cranked CHF states are added in the many-state minimization of Figure~\ref{Mg24spec}d).}
\end{figure}

To look further now into the convergence process, we present in Figure~\ref{convergence} the evolution of absolute binding energies, the energy difference with respect to the exact SM result and the transitions as a function of the number of CHF states found in the many-state minimization of Figure~\ref{Mg24spec}d). The intraband and interband transitions converge after about $\sim 5$ iterations. In binding energies, the excited state $8^+_1$ approaches the exact binding value more rapidly than the ground state. This is the effect of cranked CHF configurations and is consistent with the slight compression of the energy spectra observed in
Figure~\ref{Mg24spec}d), Figure~\ref{cranking_comparison} and Table~\ref{cranking_comparison}.

\begin{figure}[H]
  \includegraphics[scale=1.36]{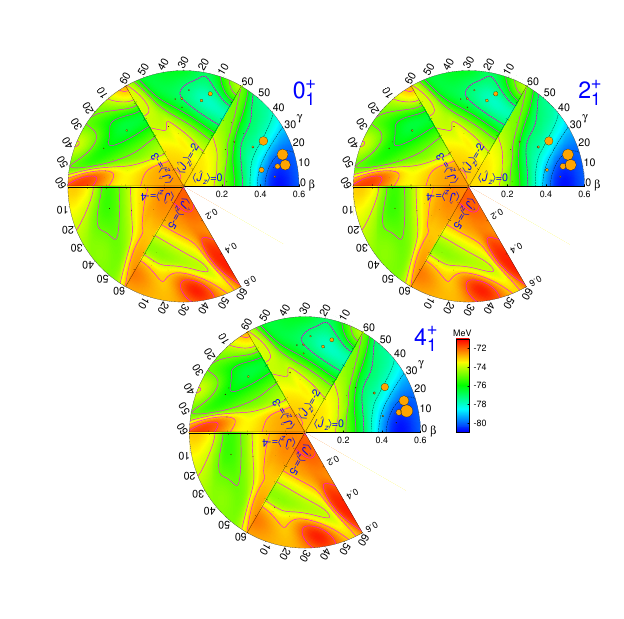}
  \caption{\label{K=0_Mg24} Structure of the first three yrast states $0^+_1$, $2^+_1$ and $4^+_1$ in the sextant $\gamma\in[0,60^{\circ}]$ of the $(\beta,\gamma)$ plane.}
\end{figure}
\begin{figure}[H] 
  \includegraphics[scale=1.36]{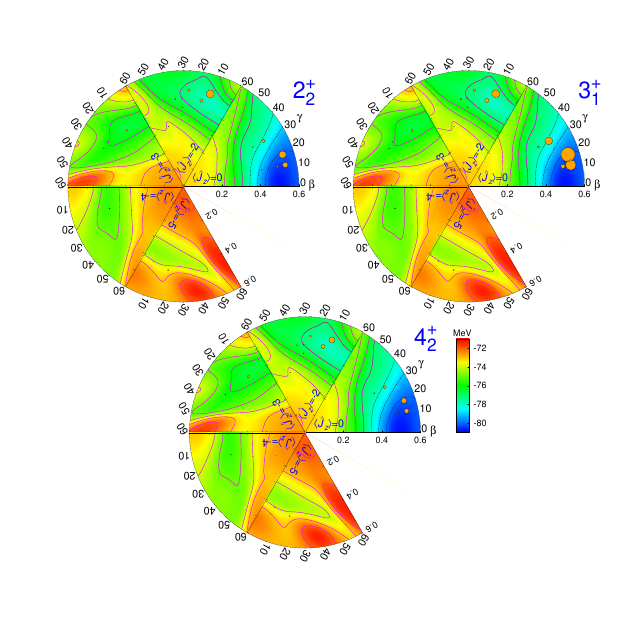}
  \caption{\label{K=2_Mg24} Structure of the first three states in the $\gamma$-band $2^+_2$, $3^+_1$ and $4^+_2$ in the sextant $\gamma\in[0,60^{\circ}]$ of the $(\beta,\gamma)$ plane.}
\end{figure}

In Figures~\ref{K=0_Mg24} and~\ref{K=2_Mg24}, the deformation structure of the yrast and the $\gamma$-band first three states is visualized in the potential energy surfaces calculated with all cranking components $\langle\hat J_z\rangle\in\{0,-2,-3,-4,-5\}$. As already shown in Figure~\ref{Mg24CHF0+1}, we see now in the full decomposition, the most important states emerge from the triaxial region around the Hartree-Fock minimum.

\begin{figure}[H]
  \scalebox{0.22}{\includegraphics{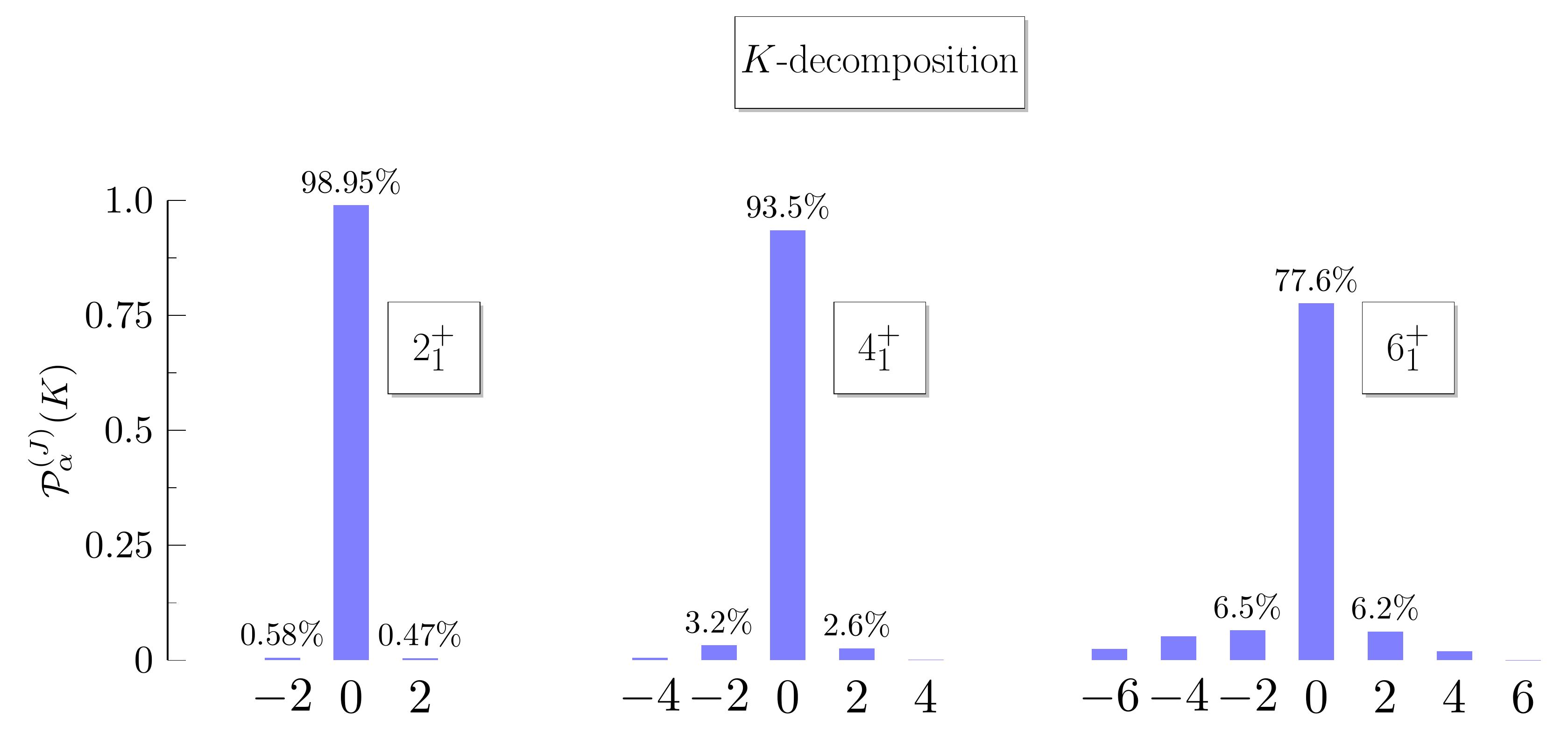}}
  \scalebox{0.22}{\includegraphics{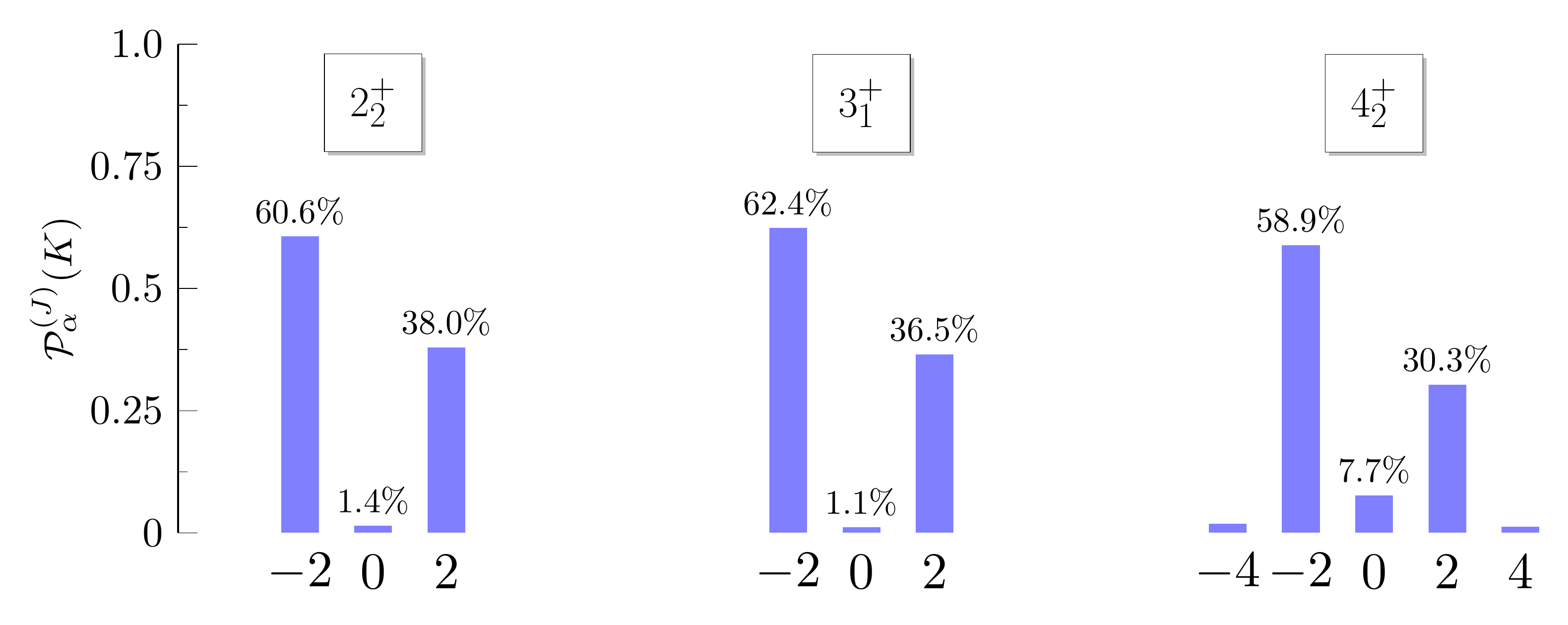}}
  \scalebox{0.22}{\includegraphics{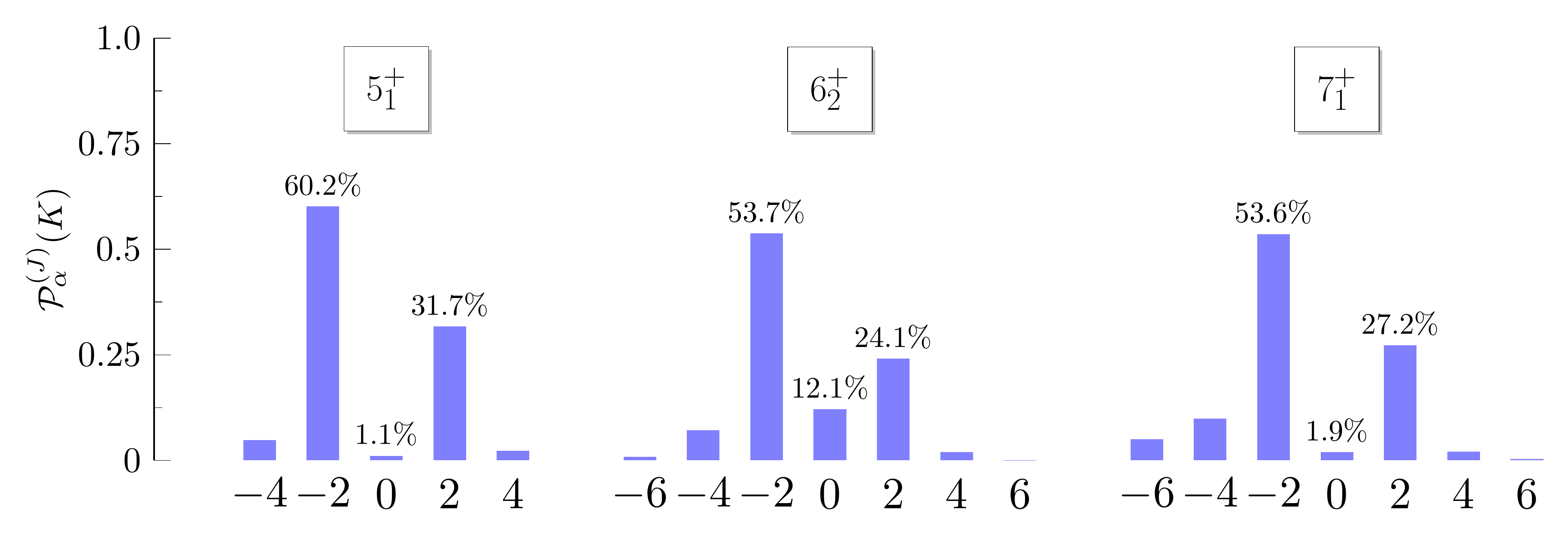}}
  \caption{\label{Mg24WF} Decomposition of states into $K$--quantum numbers in $^{24}$Mg}
\end{figure}

Physically, the triaxiality developing here is often interpreted as originating from the $K$-mixing nature of the ground state~\cite{Ring1980}. This feature can be understood from the wave function analysis in $K$-quantum numbers as presented in Figure~\ref{Mg24WF} where contributions from particular $K$-components are calculated by~\eqref{eq:prob}. Indeed, by construction single-$K$ wave functions are fully preserved by the axial contraint. The small $K$-mixing we observe in this analysis hence reflects the deviation from a pure axial picture and therefore signal the presence of triaxial configurations which we have previously analyzed. This mixing via the $K=0,2$ components also explains the weak interband transitions observed in the B$(E2)$ values.


 As a conclusion, the analysis of the wave function structures of $^{24}$Mg shows quantitatively the intimate connection between the development of triaxiality and $K$--mixing presences from a microscopic calculation using an underlying effective interaction. The ground state $K$--mixing is identified as the origin of the so-called $\gamma$-band assigned in this nucleus from geometrical models~\cite{Ring1980}. This is highly non-trivial but expected given the quality of the USDB interaction in this mass region.

\subsection{Odd nuclei: example of $^{25}$Mg}
\begin{figure}[H]
  \scalebox{0.35}{\includegraphics{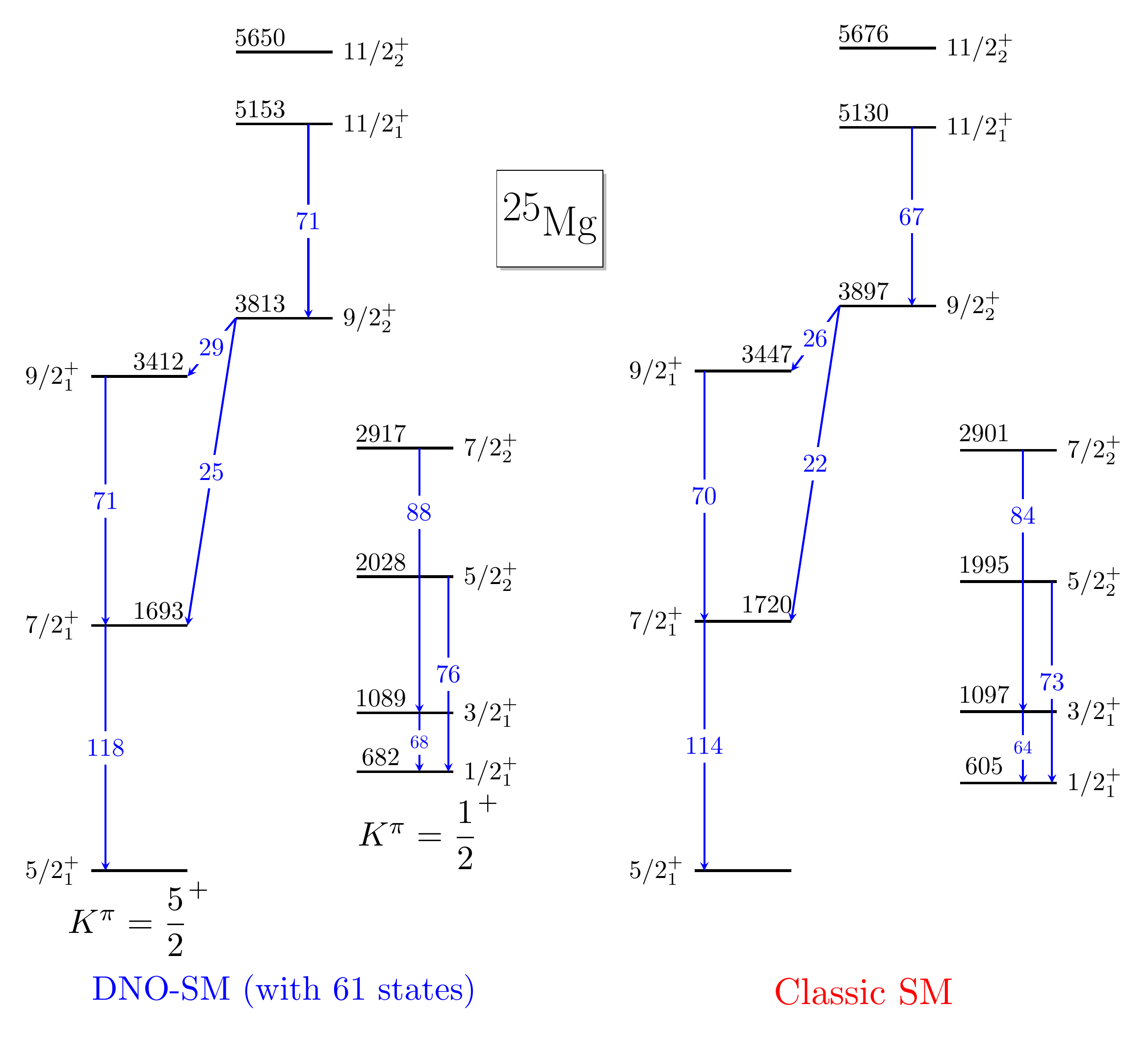}}
  \caption{\label{Mg25spec} $^{25}$Mg spectrum}
\end{figure}

\begin{figure}[H]
  \scalebox{0.17}{\includegraphics{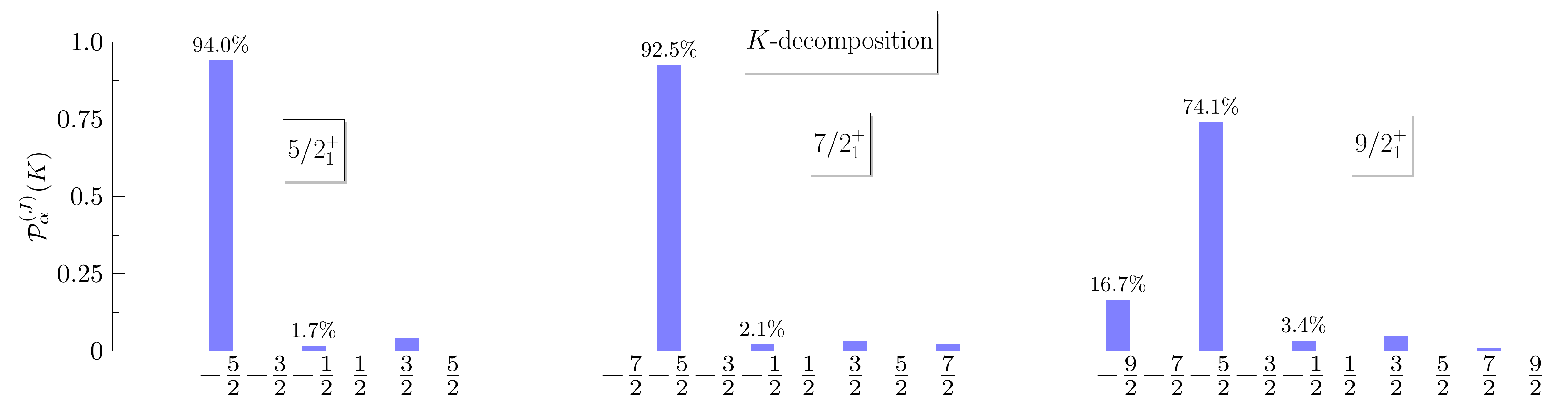}}
  \scalebox{0.18}{\includegraphics{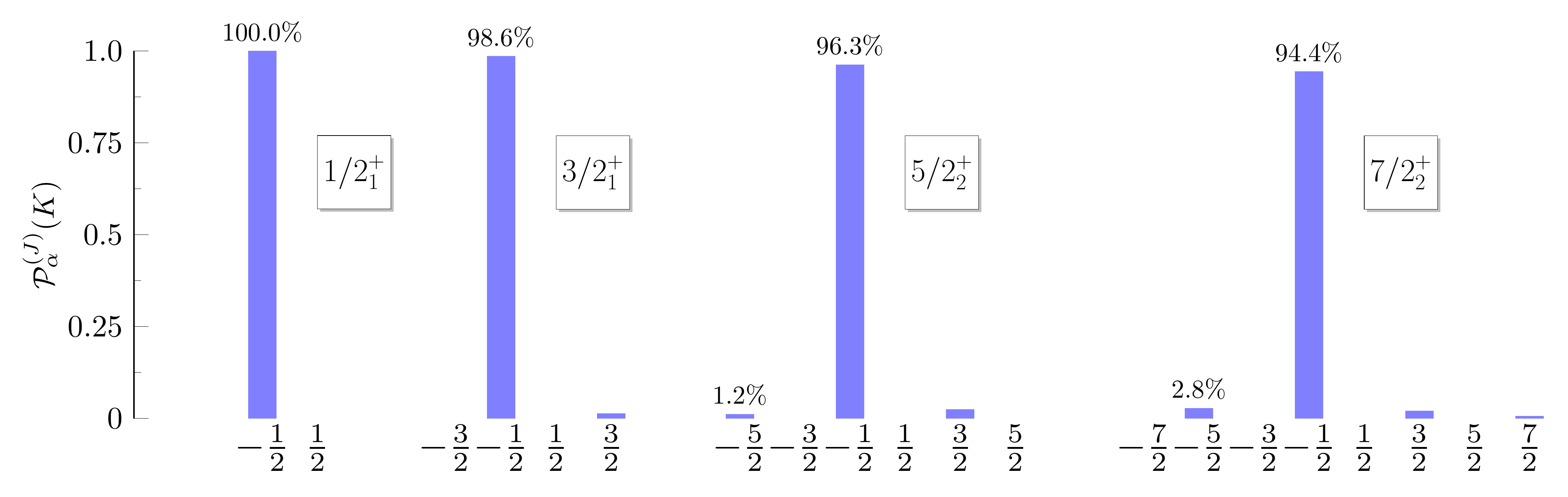}}
  \centering\includegraphics[scale=0.2]{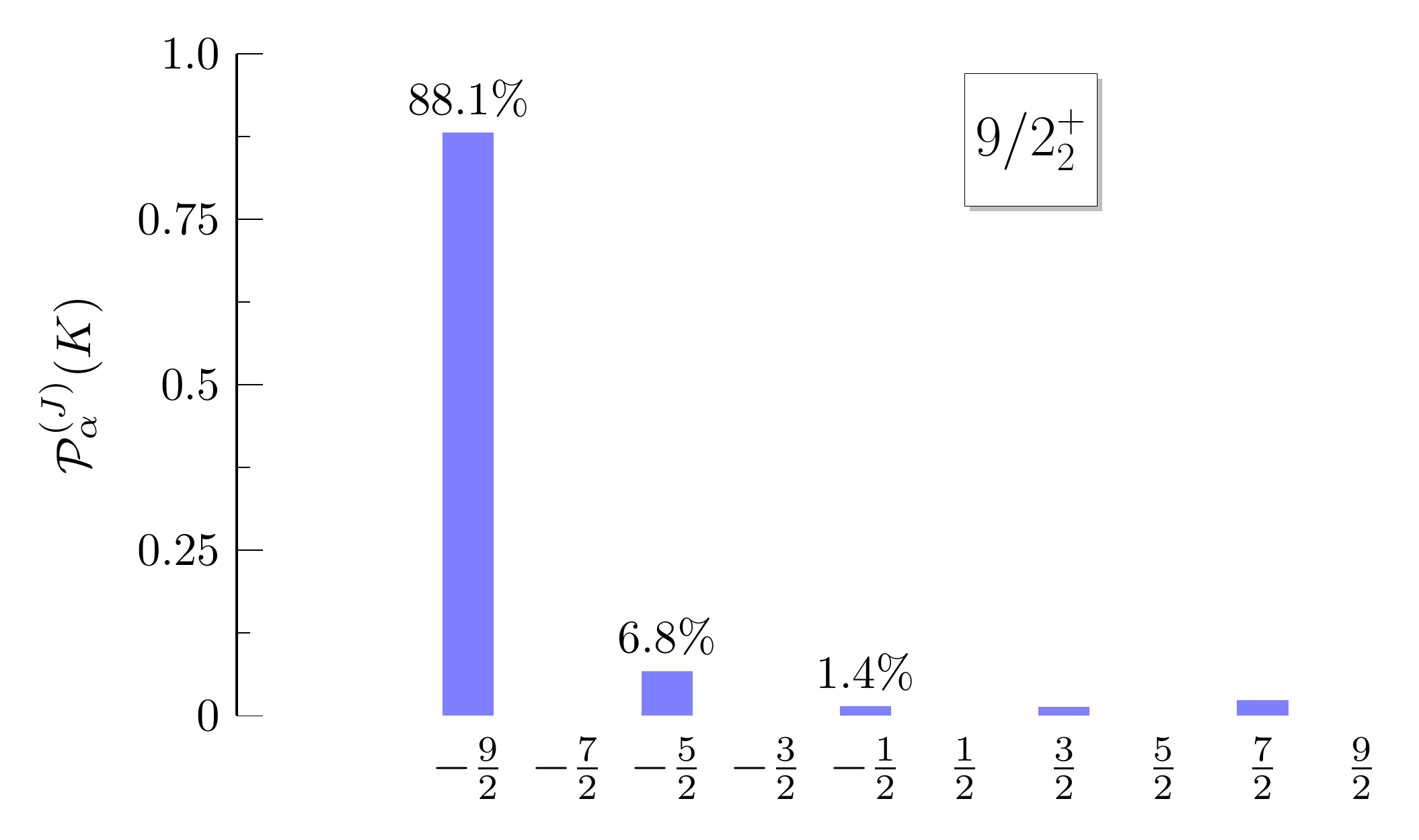}
  \scalebox{0.16}{\includegraphics{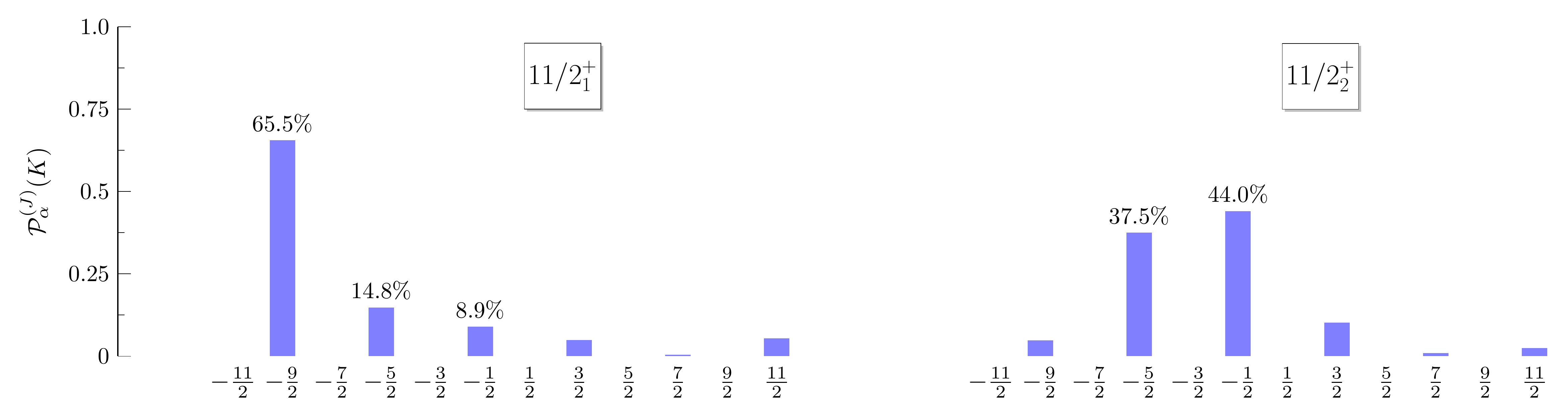}}
  \caption{\label{Mg25WF} Decomposition of states into $K$--quantum numbers in $^{25}$Mg }
\end{figure}

The above systematic benchmark has shown our DNO-SM's efficiency compared to the classic SM in even nuclei. As previously mentioned, the present framework is also capable to deal with odd nuclei in the same way without further treatments of the odd particle at the Hartree--Fock level. We find that the cranking Hartree--Fock can provide an excellent approach to set up the initial set of PCHF states from which we construct the basis $\mathit\Gamma_0$ through the diagonalization-minimization process. To illustrate this, we present a DNO-SM calculation of $^{25}$Mg whose spectrum appears in Figure~\ref{Mg25spec} compared to the classic SM one. This nucleus is known to have for instance two distinct bands: the ground state band of $K=5/2$ and an excited band of $K=1/2$. Employing the USDB effective interaction, this assignment is clearly seen in the wave function structure of each member of the bands as shown in Figure~\ref{Mg25WF}. The minimization is carried out with a discretization of $7$ points $\beta\in[0.1,0.51]$, $8$ points $\gamma\in[1.5^{\circ},60^\circ]$ and with cranking components $\langle\hat J_z\rangle=-1/2,-3/2,-5/2,-7/2,-9/2$. The spectrum of two bands is reproduced remarkably well with a rms error of $44$ keV, using $61$ CHF states. The ground state binding energy is found to be $-93.89$ MeV versus $-94.40$ MeV in the exact SM result.

In summary for the systematic comparison we have performed between the classic exact SM and our newly developed DNO-SM, we give now a brief account of its main features as well as related questions that remain to be answered

(i) Our DNO-SM formalism is composed of three elements: the existence theorem of discrete non-orthogonal bases exposed in~\eqref{eq:DNObasis}, the truncation scheme provided by the minimization technique and the diagonalization technique from the GCM theory. While the latter has been already well documented in the literature, the first element is the most important piece from the formal point of view (thanks to Ref.~\cite{Deumens1979}) because it provides us a formal language to 
express the Shell Model in a natural way with an emphasis on the different representations of the effective Hamiltonian.\\
(ii) Our minimization technique has proved to be a very efficient truncation scheme in the build-up of our non-orthogonal many-body basis for the Shell Model. In all considered nuclei, a very few number of HF states is needed to obtain solutions with very good agreement with the exact SM result.\\
(iii) The procedure constructs the full representation of the effective Hamiltonian in an iterative way where we observe that DNO-SM solutions are closer and closer from the above to the exact diagonalization limit: $\mathscr H_q \longrightarrow \mathscr H$, which means our DNO-SM is an approximation (given some particular choice of the coordinate(s) $q$) to the classic SM. This point merits to be mentioned because it points out the need to design an theoretical approach to choose the coordinate(s) $q$ if one wants to obtain the full representation of the effective Hamiltonian. Whether this question can be formally demonstrated \textit{a priori} without numerical calculations is an interesting problem to investigate.\\
(iv) The wave functions and excitation energies converge faster than the absolute energy. As depicted in Figure~\ref{Egs}, there's some missing energy in the ground state ranging around $50-900$ keV depending on specific cases. However, it has practically no impacts on the wave functions because transition probabilities are already very well reproduced (cf. Table~\ref{TAB_SD}). The physical origin of this missing binding energy is very likely related to (iii), i.e. the choice of $q$. Here we would like to mention that the multipolarity decomposition of the effective Hamiltonian designed by A. Zuker \textit{et al.} (see e.g. Ref.~\cite{RMP}) could be instrumental to investigate both (iii) and (iv).\\
(v) Lastly, the present framework is capable to treat equally well even- and odd-mass nuclei where the cranking method is found to be really efficient in the construction of the many-body basis.

To conclude, it is worth noting that the energy difference we have found using the PCHF basis is of the same order and consistent with what have been reported in similar studies of Refs.~\cite{Gao2018,Taurus,Taurus-sd,Taurus-erratum} where more sophisticated trial wave functions of Hartree-Fock-Bogoliubov (HFB) types with pairing correlations included explicitly are used (see e.g. Figure 8 of Ref.~\cite{Taurus-sd}). Finally, let us mention that our definition of $(\beta,\gamma)$ is the same as in Refs.~\cite{Taurus,Taurus-sd} up to a factor of effective charge $e^{(p)}_{mass}=e^{(n)}_{mass}$. For example, in $^{24}$Mg, omitting the effective charge would yield $\beta=0.499/1.77\approx 0.282$ which agrees well with Ref.~\cite{Taurus} (cf. Table 1 therein). We would like however to keep this effective charge factor for defining mass quadrupole moments to be able to compare with $\beta_2$--values extracted from experimental electric BE(2) transitions which yield, e.g. $0.605$ in $^{24}$Mg.

\section{A first Shell Model calculation of $^{254}\text{No}$}
\begin{figure}[H]
\begin{center}
 \scalebox{0.8}{\includegraphics{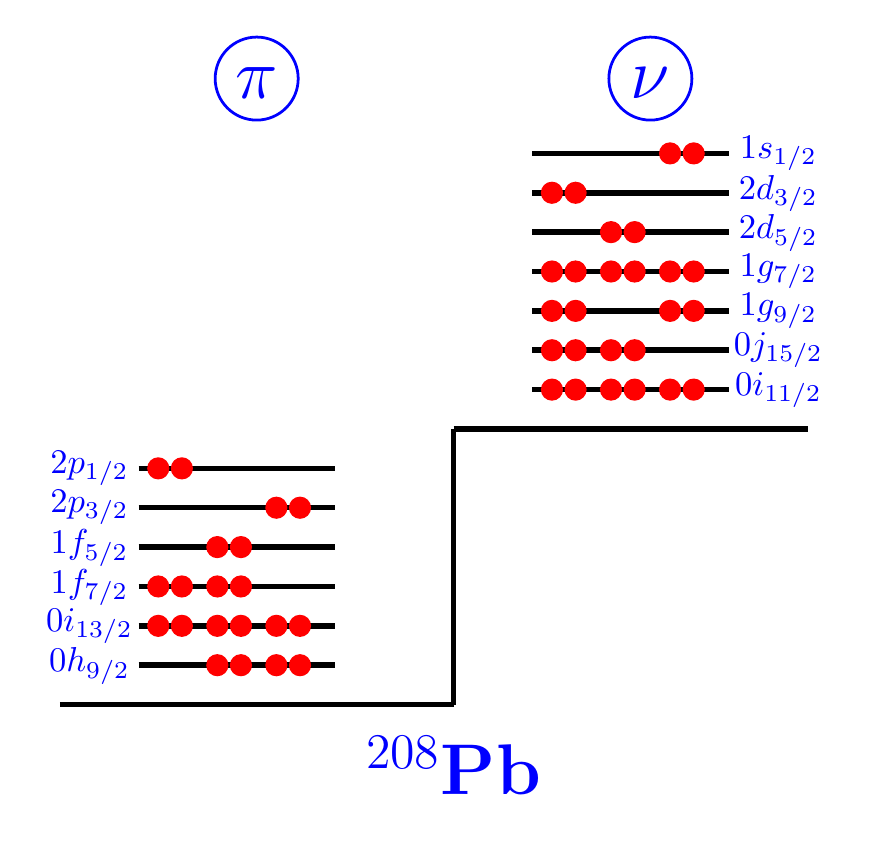}}
\end{center}
  \caption{\label{pbspace}  Valence space above $^{208}$Pb}
\end{figure}

The quest for superheavy elements is a subject  under intensive experimental investigations and usually, the predictions for  the shell stabilization of the superheavy elements predictions rely essentially on "standard" mean-field calculations. In the following, we will apply the DNO-SM method  to illustrate its applicability in the context of superheavy systems and we propose here for the first time, a shell-model type of description of $^{254}\text{No}$ superheavy nucleus, using the DNO-SM. The shell-model valence space is spanned by the full Z=82-126 proton major shell and the full N=126-184 neutron major shell beyond $^{208}\text{Pb}$, namely, the single proton orbitals  0$h_{\frac{9}{2}}$, 1$f_{\frac{7}{2}}$, 0$i_{\frac{13}{2}}$, 1$f_{\frac{5}{2}}$, 2$p_{\frac{3}{2}}$, 2$p_{\frac{1}{2}}$, and  the single neutron orbitals  1$g_{\frac{9}{2}}$, 0$i_{\frac{11}{2}}$, 0$j_{\frac{15}{2}}$, 2$d_{\frac{5}{2}}$, 3$s_{\frac{1}{2}}$, 1$g_{\frac{7}{2}}$, 2$d_{\frac{3}{2}}$
(the valence space is illustrated in in Fig.~\ref{pbspace})
. 
As  effective interaction, we use the  modified  Kuo-Herling realistic interaction (see~\cite{N=126} for details) which was applied with great success along the N=126 isotones~\cite{Th216,N=126}. The single particle energies are borrowed from $^{209}\text{Bi}$ and $^{209}\text{Pb}$ spectra for protons and neutrons respectively. The electric polarization  charge here is taken to be $\chi_p$ = $\chi_n$ = 0.5 . 

\begin{figure}[H]
  \centering 
  \includegraphics[scale=0.35]{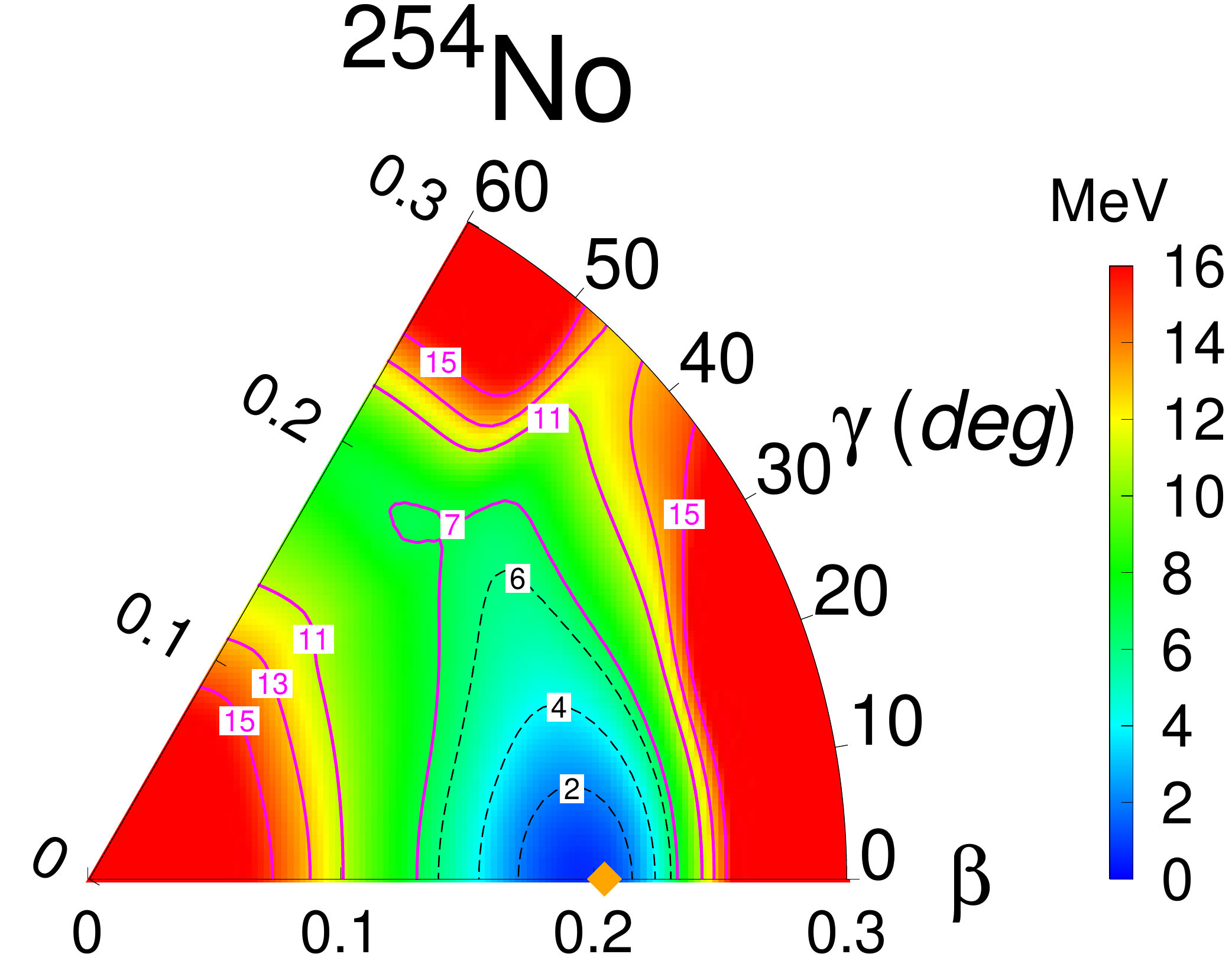}
  \caption{\label{pesNo254} Potential Energy Surface (PES) of $^{254}$No calculated with the Kuo-Herling effective interaction where the yellow diamond is the axial HF minimum with the mass quadrupole parameter $\beta = 0.2$.}
\end{figure}

$^{254}\text{No}$ is a deformed nucleus whose spectroscopy has been intensively investigated the recent years. 
In addition to the observation of its rotational Yrast structure, a side $K=3+$ band and two long lived isomers have been observed~\cite{No254-Nature,No254-PLB}. 
Figure~\ref{pesNo254} shows the potential energy surface obtained for the mass ($\beta$,$\gamma$) deformation parameters (obtained from the mass quadrupole moments defined in section~\ref{Choice of the coordinate(s) $q$}. The deformation landscape shows a clear prolate axial minimum around $\beta\sim 0.2$ extending moderately towards non-axial shapes. 

\begin{figure}[H]
  \includegraphics[scale=0.32]{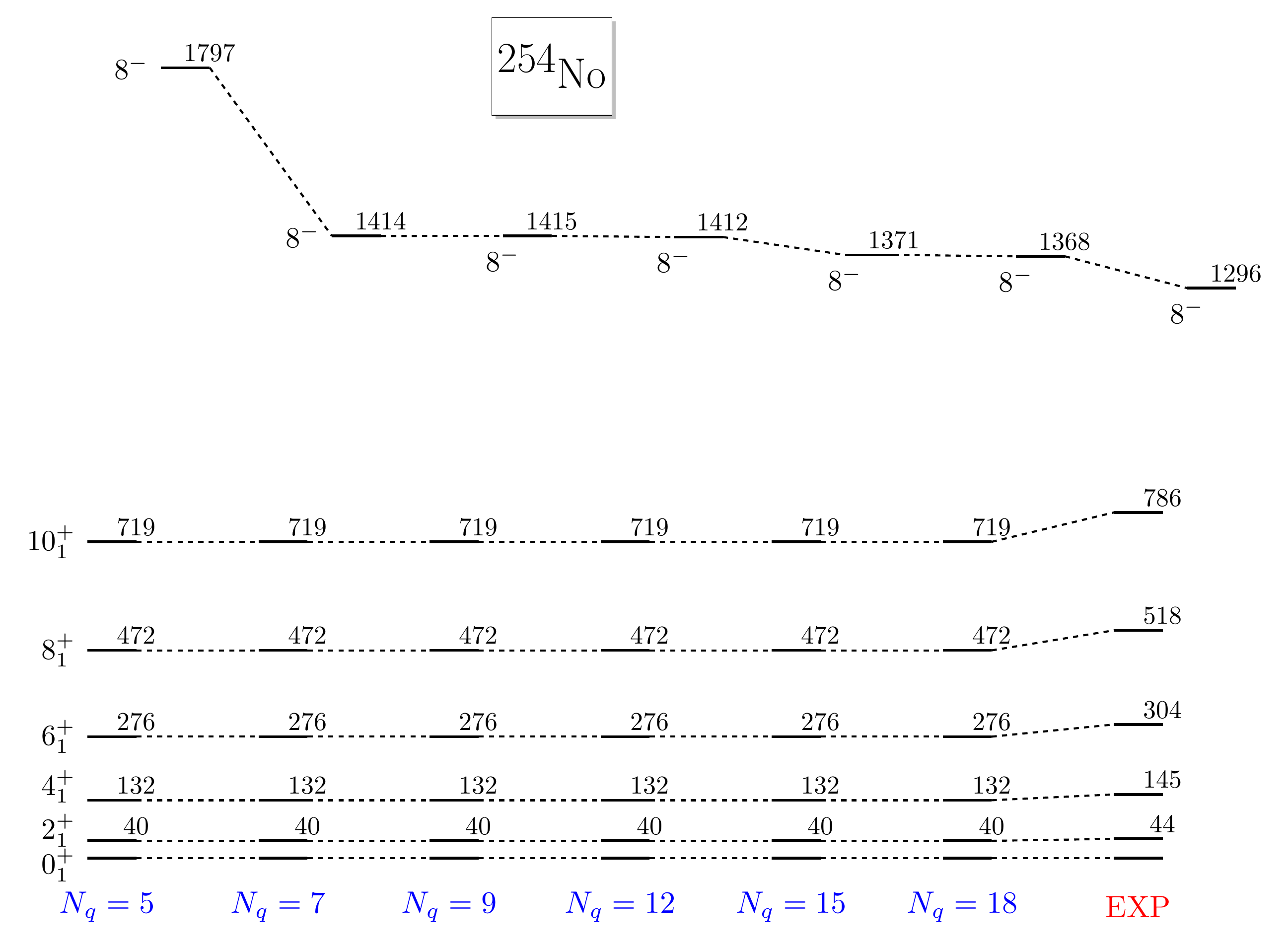}
  \caption{\label{No254spec1} Evolution of the lower part of $^{254}$No spectrum with respect to the number of HF states found by the minimization procedure.}
\end{figure}

\begin{figure}[H]
  \includegraphics[scale=0.31]{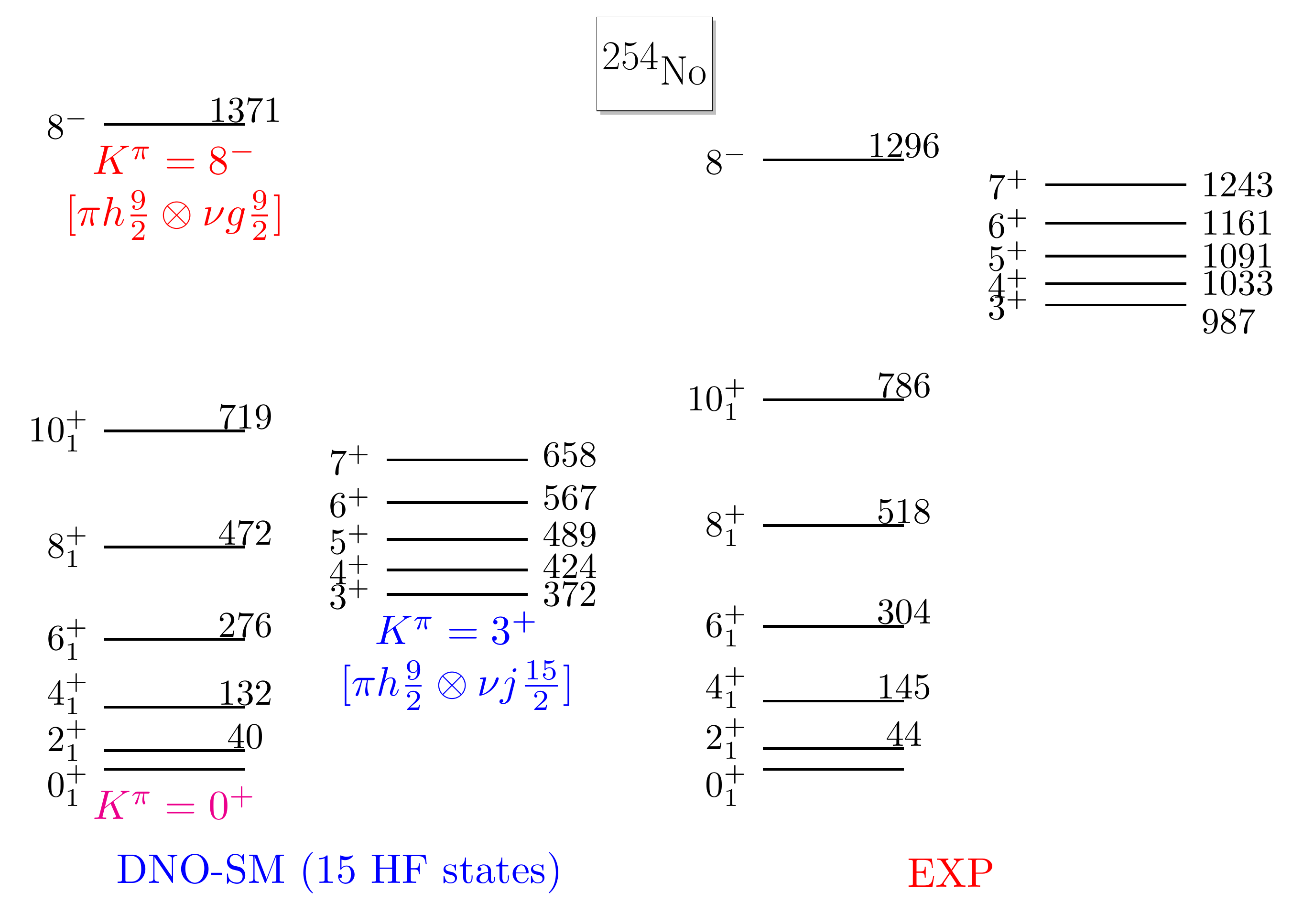}
  \caption{\label{No254spec2} Comparison of DNO-SM calculation using 15 HF states with the experimental spectrum.}
\end{figure}

All the considered states are calculated with the minimization technique over a restricted region around the corresponding HF minimum. The spectrum of $^{254}$No resulting from these calculations is shown in figure~\ref{No254spec1} as the function of the number of HF states retained by the minimization procedure. With a relatively small number of basis states, we observe a fast and good convergence of the low-lying members of the Yrast band but also for the higher lying isomer $8^-$ isomer. 
As already shown in the previous section, the DNO-SM allows analysis of the states under study in terms of intrinsic quantities, namely, deformations $(\beta,\gamma)$ and the intrinsic angular momentum. The whole low-lying spectrum is presented in figure~\ref{No254spec2}. The various states are shown and grouped in 3 structures: a $K^\pi=0^+$ Yrast rotational band, a  $K^\pi=3^+$ multiplet and the $K^\pi=8^-$ state.
There is a excellent agreement for the reproduction of the Yrast rotational sequence, and the 
$8^-$ isomeric state. The $3^-$ state banhead of the $K=3$ multiplet   is lying a little bit too low, by the interband spacing is also very well reproduced.
The formalism allows to extract the fractionnal spherical occupancies of the orbitals in the valence space. The structure of the three $0^+_1$, $3^+_1$ and $8^-_1$ states is shown in Table~\ref{occ}. with a possible large mixing of spherical orbital our description is richer than single quasi-particle estimates which are often used to assign exited and isomeric states to a single excited configurations. This is reflected in the partial occupancies of the whole proton and neutron orbitals involved in the valence space. Nevertheless, on can point out that both for protons and neutrons, the fillings proceeds through the "largest" orbitals and that  excited $3^+_1$ and $8^-_1$ states  mainly differ from the ground state by an additionnal  proton particle filling the $1h_{9/2}$ orbital, assigning these states as having a "proton" nature. We may recall here that we obtain an excellent reproduction of the experimental data with no adjustement of the effective interaction, designed more than two decades ago. Nevertheless,
in order to confirm the present outcome from our calculations, a broader systematic study has been developped, and in particular
we  would like to connect the slight energy shift of the $K^\pi=3^+$ multiplet to a specific single particle monopole drift for a
better reproductive and predictive description of the overall region.

\begin{figure}[H]
  \centering\includegraphics[scale=0.23]{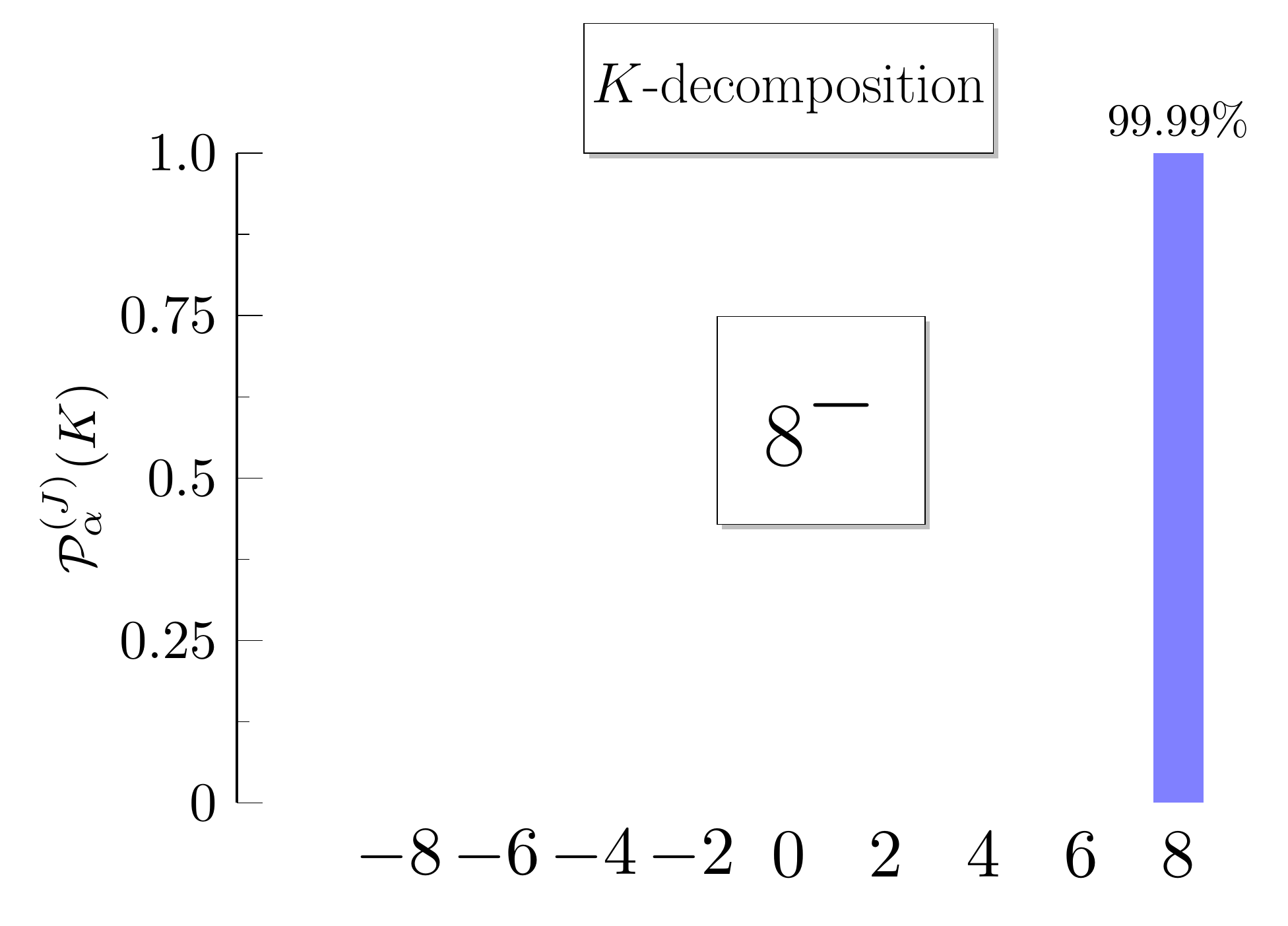}
  \includegraphics[scale=0.175]{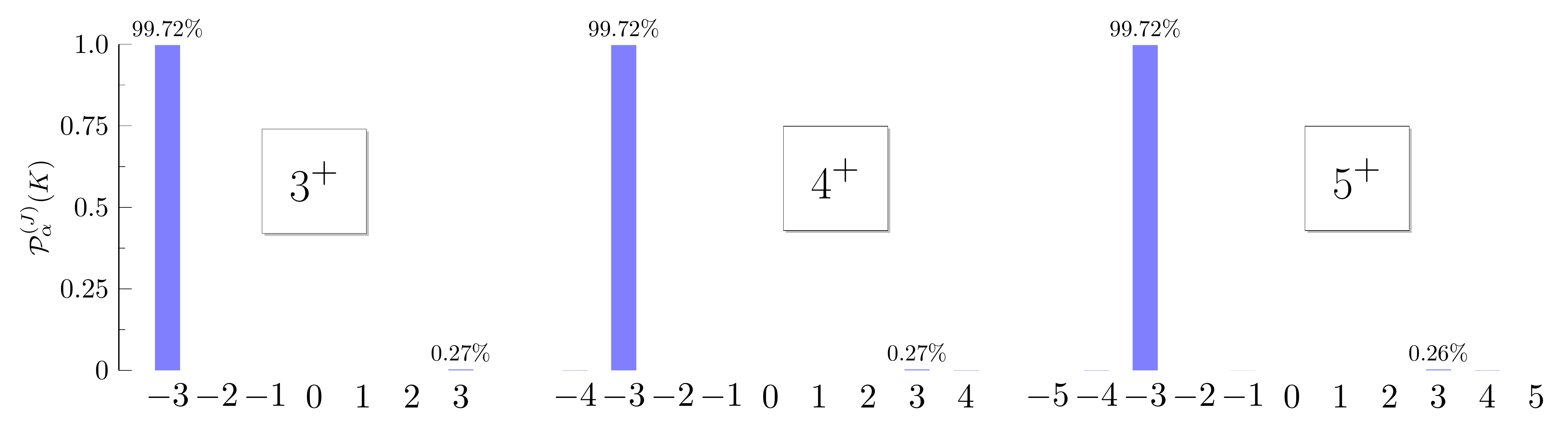}
  \includegraphics[scale=0.175]{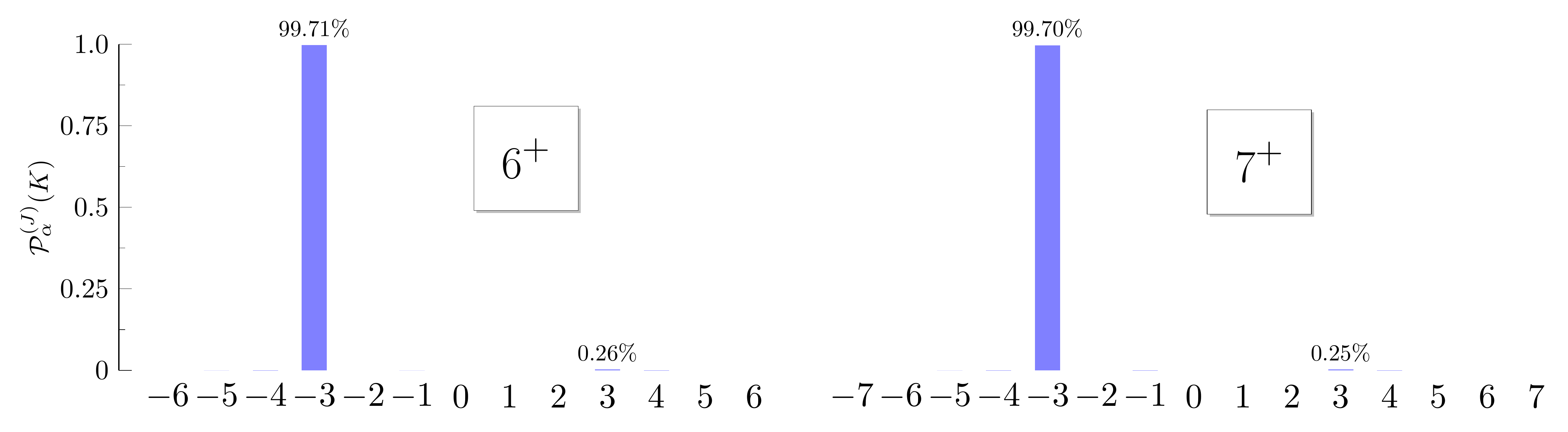}
  \caption{\label{WFNo254} $K$-quantum number content of the isomeric $3^+$ band and the $8^-$ state.}
\end{figure}

\begin{table}[H]
 \begin{tabular}{*{8}c}
    \hline\hline
    \textcolor{blue}{proton orbits} & $1h_{9/2}$ & $1i_{13/2}$ & $2f_{7/2}$ & $2f_{5/2}$ & $3p_{3/2}$ & $3p_{1/2}$ & \\
    $0^+_1$ & \textcolor{blue}{$5.66$} & \textcolor{blue}{$7.99$} & \textcolor{blue}{$3.44$} & $1.58$ & $0.76$ & $0.57$ & \\
    $8^{-}$ & \framebox{\textcolor{blue}{$6.52$}} & \textcolor{blue}{$7.82$} & \textcolor{blue}{$3.28$} & $1.20$ & $0.79$ & $0.39$ \\
    $3^{+}$ & \framebox{\textcolor{blue}{$6.50$}} & \textcolor{blue}{$7.98$} & \textcolor{blue}{$3.31$} & $1.14$ & $0.72$ & $0.35$ \\
    \hline
    \textcolor{red}{neutron orbits} & $1i_{11/2}$ & $1j_{15/2}$ & $2g_{9/2}$ & $2g_{7/2}$ & $3d_{5/2}$ & $3d_{3/2}$ & $4s_{1/2}$ \\
    $0^+_1$ & \textcolor{red}{$7.28$} & \textcolor{red}{$9.67$} & \textcolor{red}{$5.45$} & $1.11$ & $1.16$ & $0.87$ & $0.46$ \\
    $8^-$ & \textcolor{red}{$7.29$} & \framebox{\textcolor{red}{$9.04$}} & \framebox{\textcolor{red}{$6.07$}} & $1.12$ & $1.15$ & $0.88$ & $0.45$ \\
    $3^+$ & \textcolor{red}{$7.31$} & \framebox{\textcolor{red}{$9.94$}} & \textcolor{red}{$5.43$} & $0.99$ & $1.07$ & $0.83$ & $0.43$ \\
    \hline\hline
 \end{tabular}
\caption{\label{occ}Occupancies of  the spherical  orbitals for the ground state and  $8^-$ and $3^+$ states.}
\end{table}

\section{Conclusion and perspectives}
As a summary, in this paper we have exposed the formalism of the DNO-SM which amounts to diagonalize shell-model hamiltonians in a non-orthogonal basis with the use of beyond-mean-field techniques. Particular effort has been put into the proper selection of optimal basis state used for the diagonalisation.
We benchmarked the method over a large set of $sd$ shell nuclei 
and could reproduce the energies, and transitions probabilities
of the exact diagonalisations with minimal cost of very few basis states. For the first time, we applied the DNO-SM method to a superheavy system $^{254}$No which is obviously far beyond the capabilities of actual diagonalisations codes.
The DNO-SM formalism has also recently been very useful in the interpretation of several experimental studies~\cite{Zn70,Arsenics,Bromes,Zn74} and has been to be extremely promising both for instrinsic interpretation of shell-model diagonalisations, and setting new frontiers for nuclear structure studies within the shell-model framework.

\section{Acknowlegments} The authors would like to dedicate the present work to the memory of the  late Etienne Caurier, who was  at the initiative of the present developments.
\appendix

\section{Matrix elements in the PCHF basis}

Considering the two-body Hamiltonian $\mathcal{\hat H}$ defined in~\eqref{eq:hamiltonian}, in the PCHF basis $\mathcal P^J_{MK}\ket{\Phi(q)}\in\mathit\Gamma_0$, it is represented by the set of matrix elements $\elmx{\Phi(q')}{\mathcal{\hat H}\mathcal P^J_{K'K}}{\Phi(q)}$ given by~\eqref{eq:TBME_PCHF}. The calculation of this matrix element requires an evaluation of the hamiltonian kernel which, for two arbitrary Slater determinants $\ket{\Phi'}$ and $\ket\Phi$, takes the forms (cf. e.g. Ref.~\cite{Watt1972})
\begin{widetext}
\begin{equation}
  \elmx{\Phi'}{\mathcal{\hat H} \hat R(\Omega)}{\Phi} =
  \sum_{\substack{p\in\Phi'\\q\in\Phi}}(-)^{p+q} M_{pq}(\Omega)\,\elmx{p}{\hat E \hat R(\Omega)}{q}
  + \sum_{\substack{p<q\in\Phi'\\r<s\in\Phi}}(-)^{p+q+r+s}M_{pqrs}(\Omega)
  \elmx{pq}{\hat{\mathcal V} \hat R(\Omega)}{rs}
\end{equation}
\end{widetext}
where $M_{pq}(\Omega),M_{pqrs}(\Omega)$ are respectively first-- and second--order minors of the $A\times A$ matrix $N(\Omega) = D^{\prime\dagger}\cdot R(\Omega)\cdot D$ with a rectangular matrix $D_{ip} = \{C^{(p)}_i,\,p\in\Phi,i\in\mathcal E\}$ representing the single particle HF states in the Slater $\ket\Phi$. $C^{(p)}_i$ represents here the expansion coefficient of the single-particle Hartree-Fock state $\ket{p}$ in the single-particle Harmonic Oscillator basis $\{\ket{i}\}$. The matrix element of the one-body single-particle energy $\hat E$ is given by
\begin{equation}
  \elmx{p}{\hat E \hat R(\Omega)}{q} = \sum_{i_1,i_2,i\in\mathcal E}
  C^{\prime (p)}_{i_1}C^{(q)}_{i_2}\,e_{i_1i}\, R_{ii_2}(\Omega).
\end{equation}
Whereas for the two-body term, it is written in terms of the antisymmetrized matrix element $\elmx{JT(i_1i_2)}{\hat{\mathcal V}}{JT(i_3i_4)}$ of good angular momentum and isospin $J,T$
\begin{widetext}
\begin{equation}
  \elmx{pq}{\hat{\mathcal V} \hat R(\Omega)}{rs} = \sum_{\substack{JMM'\\TT_z\\i_1i_2i_3i_4}}
    C^{TT_z}_{\frac 1 2\tau_p\frac 1 2\tau_q}C^{TT_z}_{\frac 1 2\tau_r\frac 1 2\tau_s}
    C^{JM}_{j_1m_1j_2m_2}C^{JM'}_{j_3m_3j_4m_4}
    \mathcal D^J_{MM'}(\Omega) \,
    C^{(p)}_{i_1}C^{(q)}_{i_2}C^{(r)}_{i_3}C^{(s)}_{i_4} \: \elmx{JT(i_1i_2)}{\hat{\mathcal V}}{JT(i_3i_4)}.
\end{equation}
\end{widetext}
Here $C^{JM}_{j_1m_1j_2m_2}$ is the Clebsch--Gordan coefficient.

\section{Derivation of the analytical integration over the Euler angles $\alpha,\gamma$}

Appendix A provides the necessary elements we need to derive an analytical formula for the integrations over $\alpha,\gamma$ in~\eqref{eq:TBME_PCHF}. For that goal, we just need to perform the derivation with $\hat{\mathcal O} = \mathbf 1$, the same procedure holds for other operators such as the Hamiltonian or transition operators. In this case, we have the norm matrix element 
\begin{equation}
  \begin{aligned}
    \mathcal N^{J}_{K'K} = \frac{2J+1}{4\pi^2\big(3-(-)^A\big)}
    \int d\Omega \: \mathcal D^{J*}_{MK}(\Omega) \: \mathrm{det}\,N(\Omega)
  \end{aligned}
  \label{eq:norm_matrix_element}
\end{equation}
where we have used the equality $\mathrm{det}\,N(\Omega) = \elmx{\Phi'}{\hat R(\Omega)}{\Phi}$. Let us rewrite this quantity in an explicit way with a summation over all permutations $\sigma$ of the permutation group of $A$--particles $S_A$
\begin{equation}
  \begin{aligned}
    &\mathrm{det}\,N = \sum_{\sigma\in S_A} \mathrm{sgn}(\sigma) \prod_{\lambda=1}^A N_{\lambda,\sigma(\lambda)} \\
    &= \sum_{\sigma\in S_A} \mathrm{sgn}(\sigma) \prod_{\lambda=1}^A
    \Big(\sum_{i_1i_2\in\mathcal E}D'_{\lambda i_1} R_{i_1i_2} D_{i_2\sigma(\lambda)}\Big) \\
    &= \sum_{\substack{i_1(1)i_2(1)\\\cdots\\i_1(A)i_2(A)}} \prod_{\lambda=1}^A D^{\prime\dagger}_{\lambda i_1(\lambda)}
    \prod_{\lambda=1}^A R_{i_1(\lambda)i_2(\lambda)} \\
    &\phantom{=}\Big(\sum_{\sigma\in S_A} \mathrm{sgn}(\sigma)\prod_{\lambda=1}^AD_{i_2(\lambda)\sigma(\lambda)}\Big).
  \end{aligned}
\end{equation}
The norm matrix element thus becomes
\begin{equation}
  \begin{aligned}
    &\mathcal N^{(J)}_{K'K} = \frac{2J+1}{4\pi^2\big(3-(-)^A\big)} 
    \sum_{\substack{i_1(1)i_2(1)\\\cdots\\i_1(A)i_2(A)}} \\
    &\prod_{\lambda=1}^A D^{\prime\dagger}_{\lambda i_1(\lambda)} 
    \Big(\sum_{\sigma\in S_A} \mathrm{sgn}(\sigma)\prod_{\lambda=1}^AD_{i_2(\lambda)\sigma(\lambda)}\Big) \times \\
    &\int d\Omega\: \mathcal D^{J*}_{MK}(\Omega) \prod_{\lambda=1}^A R_{i_1(\lambda)i_2(\lambda)}(\Omega).
  \end{aligned}
\end{equation}
In this form, the integration over Euler angles $\Omega=(\alpha,\beta,\gamma)$ is isolated and can be subject to a direct evaluation using the rotation matrix in spherical oscillator basis $R_{i_1i_2}(\Omega) = e^{-i\alpha m_1}d^j_{m_1m_2}(\beta)e^{-i\gamma m_2}$. Hence, one can write
\begin{equation}
  \begin{aligned}
    &\int d\Omega\: \mathcal D^{J*}_{MK}(\Omega) \prod_{\lambda=1}^A R_{i_1(\lambda)i_2(\lambda)}(\Omega) \\
    &= \int\displaylimits_0^{2\pi}d\alpha \, e^{i\alpha\big[K'-\sum_{\lambda=1}^{A} m_1(\lambda)\big]} 
    \int\displaylimits_{0}^{\gamma_{\rm max}}d\gamma\, e^{i\gamma\big[K-\sum_{\lambda=1}^{A} m_2(\lambda)\big]} \\
    &\phantom{=}\int\displaylimits_0^{\pi}d\beta\,\sin\beta d^{*J}_{K'K}(\beta) \prod_{\lambda=1}^A d^{j(\lambda)}_{m_1(\lambda)m_2(\lambda)}(\beta)
  \end{aligned}
\end{equation}
from which it is trivial to calculate the integrals over $\alpha,\gamma$. The result simply reads
\begin{equation}
  \begin{aligned}
    &\int d\Omega\: \mathcal D^{J*}_{MK}(\Omega) \prod_{\lambda=1}^A R_{i_1(\lambda)i_2(\lambda)}(\Omega)
    = 2\pi\gamma_{\rm max} \times \\
    &\delta_{\Delta K',0}\delta_{\Delta K,0}
    \int\displaylimits_0^\pi d\beta\sin\beta\,
    d^{*J}_{K'K}(\beta) \prod_{\lambda=1}^A d^{j(\lambda)}_{m_1(\lambda)m_2(\lambda)}(\beta)
  \end{aligned}
\end{equation}
where we denote $\Delta K' = K' - \sum_{\lambda=1}^A m_1(\lambda)$ and $\Delta K = K - \sum_{\lambda=1}^A m_2(\lambda)$. This form is of course not practical as it involves summations over $A!$ permutations. Our next step is to recast it into the familiar expression as in~\eqref{eq:norm_matrix_element} with only the integration over $\beta$ being left to evaluate. To do so, notice that
\begin{equation}
  \delta_{n,0} = \frac 1 N \sum_{k=1}^N e^{i\frac{2\pi k}{N}n}
  \text{ if }\left\{\begin{aligned}
  N\in\mathbb N, \: N \geq 2, \:\:\\
  n\in\mathbb Z, \: e^{i\frac{2\pi}{N}n} \neq 1.
  \end{aligned}\right.
  \label{eq:delta_condition}
\end{equation}
It is obvious that $\Delta K', \Delta K \in\mathbb Z$ regardless the odd or even mass number $A$ so that applying this identity allows us to obtain
\begin{equation}
  \begin{aligned}
    &\int d\Omega\: \mathcal D^{J*}_{MK}(\Omega) \prod_{\lambda=1}^A R_{i_1(\lambda)i_2(\lambda)}(\Omega)
    = \frac{2\pi\gamma_{\rm max}}{N_\alpha N_\gamma} \times \\
    &\sum_{k_1=1}^{N_\alpha}\sum_{k_2=1}^{N_\gamma}
    \int\displaylimits_0^\pi d\beta\sin\beta\,\mathcal D^{*J}_{K'K}\big(\frac{2\pi k_1}{N_\alpha},\beta,\frac{2\pi k_2}{N_\gamma}\big) \times \\
    &\prod_{\lambda=1}^A R_{i_1(\lambda)i_2(\lambda)}\big(\frac{2\pi k_1}{N_\alpha},\beta,\frac{2\pi k_2}{N_\gamma}\big)
  \end{aligned}
\end{equation}
where $N_\alpha,N_\gamma\in\mathbb Z$ are chosen according to the condition~\eqref{eq:delta_condition}. The final expression of the norm matrix element thus reads
\begin{equation}
  \begin{aligned}
    \mathcal N^J_{K'K} = \frac{2J+1}{2}\cdot\frac{1}{N_\alpha N_\gamma} \sum_{k_1=1}^{N_\alpha}\sum_{k_2=1}^{N_\gamma}
    \int\displaylimits_0^\pi d\beta\sin\beta \:\times \\
    \mathcal D^{*J}_{K'K}\big(\frac{2\pi k_1}{N_\alpha},\beta,\frac{2\pi k_2}{N_\gamma}\big) \times
    \mathrm{det}\,N\big(\frac{2\pi k_1}{N_\alpha},\beta,\frac{2\pi k_2}{N_\gamma}\big).
  \end{aligned}
\end{equation}
The same reasoning leaves us with the Hamiltonian matrix element where $\alpha,\gamma$ are exactly integrated out
\begin{widetext}
\begin{equation}
  \begin{aligned}
    \mathcal H^J_{K'K} = \frac{2J+1}{2}\cdot\frac{1}{N_\alpha N_\gamma} \sum_{k_1=1}^{N_\alpha}\sum_{k_2=1}^{N_\gamma}
    \int\displaylimits_0^\pi d\beta\sin\beta \:
    \mathcal D^{*J}_{K'K}\big(\frac{2\pi k_1}{N_\alpha},\beta,\frac{2\pi k_2}{N_\gamma}\big) \:
    \elmx{\Phi'}{\hat{\mathcal H}\hat R\big(\frac{2\pi k_1}{N_\alpha},\beta,\frac{2\pi k_2}{N_\gamma}\big)}{\Phi}.
  \end{aligned}
\end{equation}
\end{widetext}
$N_u (u=\alpha,\gamma)$ will be chosen to ensure the conditions
\begin{widetext}
\begin{equation}
  e^{i\frac{2\pi}{N_u}\Delta K_v} \neq 1 \quad \forall v=0,1,2 \text{  with  }
  \left\{\begin{aligned}
  &\Delta K_0 = K - \sum_{\lambda=1}^A m_i(\lambda) \quad \forall i\in\mathcal E \\
  &\Delta K_1 = K - m_i - \sum_{\lambda=1}^{A-1} m_{i'}(\lambda) \quad \forall i,i'\in\mathcal E \\
  &\Delta K_2 = K - M(i_1,i_2) - \sum_{\lambda=1}^{A-2} m_i(\lambda) \quad \forall i,i_1,i_2\in\mathcal E \\
  \end{aligned}\right.
\end{equation}
\end{widetext}
with $i_1,i_2$ designating two single-particle harmonics oscillator states coupled to a total angular momentum $M=m_{i_1}+m_{i_2}$.

\bibliography{paper}

\begin{thebibliography}{62}
\expandafter\ifx\csname natexlab\endcsname\relax\def\natexlab#1{#1}\fi
\expandafter\ifx\csname bibnamefont\endcsname\relax
  \def\bibnamefont#1{#1}\fi
\expandafter\ifx\csname bibfnamefont\endcsname\relax
  \def\bibfnamefont#1{#1}\fi
\expandafter\ifx\csname citenamefont\endcsname\relax
  \def\citenamefont#1{#1}\fi
\expandafter\ifx\csname url\endcsname\relax
  \def\url#1{\texttt{#1}}\fi
\expandafter\ifx\csname urlprefix\endcsname\relax\def\urlprefix{URL }\fi
\providecommand{\bibinfo}[2]{#2}
\providecommand{\eprint}[2][]{\url{#2}}

\bibitem[{\citenamefont{Caurier et~al.}(2005)\citenamefont{Caurier,
  Mart\'{\i}nez-Pinedo, Nowacki, Poves, and Zuker}}]{RMP}
\bibinfo{author}{\bibfnamefont{E.}~\bibnamefont{Caurier}},
  \bibinfo{author}{\bibfnamefont{G.}~\bibnamefont{Mart\'{\i}nez-Pinedo}},
  \bibinfo{author}{\bibfnamefont{F.}~\bibnamefont{Nowacki}},
  \bibinfo{author}{\bibfnamefont{A.}~\bibnamefont{Poves}}, \bibnamefont{and}
  \bibinfo{author}{\bibfnamefont{A.~P.} \bibnamefont{Zuker}},
  \bibinfo{journal}{Rev. Mod. Phys.} \textbf{\bibinfo{volume}{77}},
  \bibinfo{pages}{427} (\bibinfo{year}{2005}),
  \urlprefix\url{https://link.aps.org/doi/10.1103/RevModPhys.77.427}.

\bibitem[{\citenamefont{Nowacki et~al.}(2021)\citenamefont{Nowacki, Obertelli,
  and Poves}}]{PPNP}
\bibinfo{author}{\bibfnamefont{F.}~\bibnamefont{Nowacki}},
  \bibinfo{author}{\bibfnamefont{A.}~\bibnamefont{Obertelli}},
  \bibnamefont{and} \bibinfo{author}{\bibfnamefont{A.}~\bibnamefont{Poves}},
  \bibinfo{journal}{Progress in Particle and Nuclear Physics}
  \textbf{\bibinfo{volume}{120}}, \bibinfo{pages}{103866}
  (\bibinfo{year}{2021}), ISSN \bibinfo{issn}{0146-6410},
  \urlprefix\url{https://www.sciencedirect.com/science/article/pii/S014664102100020X}.

\bibitem[{\citenamefont{Caurier et~al.}(1999)\citenamefont{Caurier,
  Mart\'{\i}nez-Pinedo, Nowacki, Poves, Retamosa, and Zuker}}]{Caurier:1998zw}
\bibinfo{author}{\bibfnamefont{E.}~\bibnamefont{Caurier}},
  \bibinfo{author}{\bibfnamefont{G.}~\bibnamefont{Mart\'{\i}nez-Pinedo}},
  \bibinfo{author}{\bibfnamefont{F.}~\bibnamefont{Nowacki}},
  \bibinfo{author}{\bibfnamefont{A.}~\bibnamefont{Poves}},
  \bibinfo{author}{\bibfnamefont{J.}~\bibnamefont{Retamosa}}, \bibnamefont{and}
  \bibinfo{author}{\bibfnamefont{A.~P.} \bibnamefont{Zuker}},
  \bibinfo{journal}{Phys. Rev. C} \textbf{\bibinfo{volume}{59}},
  \bibinfo{pages}{2033} (\bibinfo{year}{1999}),
  \urlprefix\url{https://link.aps.org/doi/10.1103/PhysRevC.59.2033}.

\bibitem[{\citenamefont{Nowacki et~al.}(2016)\citenamefont{Nowacki, Poves,
  Caurier, and Bounthong}}]{Nowacki:2016isq}
\bibinfo{author}{\bibfnamefont{F.}~\bibnamefont{Nowacki}},
  \bibinfo{author}{\bibfnamefont{A.}~\bibnamefont{Poves}},
  \bibinfo{author}{\bibfnamefont{E.}~\bibnamefont{Caurier}}, \bibnamefont{and}
  \bibinfo{author}{\bibfnamefont{B.}~\bibnamefont{Bounthong}},
  \bibinfo{journal}{Phys. Rev. Lett.} \textbf{\bibinfo{volume}{117}},
  \bibinfo{pages}{272501} (\bibinfo{year}{2016}),
  \urlprefix\url{https://link.aps.org/doi/10.1103/PhysRevLett.117.272501}.

\bibitem[{\citenamefont{Hinke et~al.}(2012)\citenamefont{Hinke, Böhmer,
  Boutachkov et~al.}}]{Sn100}
\bibinfo{author}{\bibfnamefont{C.}~\bibnamefont{Hinke}},
  \bibinfo{author}{\bibfnamefont{M.}~\bibnamefont{Böhmer}},
  \bibinfo{author}{\bibfnamefont{P.}~\bibnamefont{Boutachkov}},
  \bibnamefont{et~al.}, \bibinfo{journal}{Nature}
  \textbf{\bibinfo{volume}{341}}, \bibinfo{pages}{341} (\bibinfo{year}{2012}).

\bibitem[{\citenamefont{Siciliano et~al.}(2020)\citenamefont{Siciliano,
  Valiente-Dobón, Goasduff, Nowacki, Zuker, Bazzacco, Lopez-Martens, Clément,
  Benzoni, Braunroth et~al.}}]{Siciliano:2019qhw}
\bibinfo{author}{\bibfnamefont{M.}~\bibnamefont{Siciliano}},
  \bibinfo{author}{\bibfnamefont{J.}~\bibnamefont{Valiente-Dobón}},
  \bibinfo{author}{\bibfnamefont{A.}~\bibnamefont{Goasduff}},
  \bibinfo{author}{\bibfnamefont{F.}~\bibnamefont{Nowacki}},
  \bibinfo{author}{\bibfnamefont{A.}~\bibnamefont{Zuker}},
  \bibinfo{author}{\bibfnamefont{D.}~\bibnamefont{Bazzacco}},
  \bibinfo{author}{\bibfnamefont{A.}~\bibnamefont{Lopez-Martens}},
  \bibinfo{author}{\bibfnamefont{E.}~\bibnamefont{Clément}},
  \bibinfo{author}{\bibfnamefont{G.}~\bibnamefont{Benzoni}},
  \bibinfo{author}{\bibfnamefont{T.}~\bibnamefont{Braunroth}},
  \bibnamefont{et~al.}, \bibinfo{journal}{Physics Letters B}
  \textbf{\bibinfo{volume}{806}}, \bibinfo{pages}{135474}
  (\bibinfo{year}{2020}), ISSN \bibinfo{issn}{0370-2693},
  \urlprefix\url{https://www.sciencedirect.com/science/article/pii/S0370269320302781}.

\bibitem[{\citenamefont{Rosiak et~al.}(2018)\citenamefont{Rosiak, Seidlitz,
  Reiter, Na\"{\i}dja, Tsunoda, Togashi, Nowacki, Otsuka, Col\`o, Arnswald
  et~al.}}]{MINIBALL:2018zvw}
\bibinfo{author}{\bibfnamefont{D.}~\bibnamefont{Rosiak}},
  \bibinfo{author}{\bibfnamefont{M.}~\bibnamefont{Seidlitz}},
  \bibinfo{author}{\bibfnamefont{P.}~\bibnamefont{Reiter}},
  \bibinfo{author}{\bibfnamefont{H.}~\bibnamefont{Na\"{\i}dja}},
  \bibinfo{author}{\bibfnamefont{Y.}~\bibnamefont{Tsunoda}},
  \bibinfo{author}{\bibfnamefont{T.}~\bibnamefont{Togashi}},
  \bibinfo{author}{\bibfnamefont{F.}~\bibnamefont{Nowacki}},
  \bibinfo{author}{\bibfnamefont{T.}~\bibnamefont{Otsuka}},
  \bibinfo{author}{\bibfnamefont{G.}~\bibnamefont{Col\`o}},
  \bibinfo{author}{\bibfnamefont{K.}~\bibnamefont{Arnswald}},
  \bibnamefont{et~al.} (\bibinfo{collaboration}{MINIBALL and HIE-ISOLDE
  Collaborations}), \bibinfo{journal}{Phys. Rev. Lett.}
  \textbf{\bibinfo{volume}{121}}, \bibinfo{pages}{252501}
  (\bibinfo{year}{2018}),
  \urlprefix\url{https://link.aps.org/doi/10.1103/PhysRevLett.121.252501}.

\bibitem[{\citenamefont{Na\"{\i}dja et~al.}(2017)\citenamefont{Na\"{\i}dja,
  Nowacki, and Bounthong}}]{Naidja:2017tyv}
\bibinfo{author}{\bibfnamefont{H.}~\bibnamefont{Na\"{\i}dja}},
  \bibinfo{author}{\bibfnamefont{F.}~\bibnamefont{Nowacki}}, \bibnamefont{and}
  \bibinfo{author}{\bibfnamefont{B.}~\bibnamefont{Bounthong}},
  \bibinfo{journal}{Phys. Rev. C} \textbf{\bibinfo{volume}{96}},
  \bibinfo{pages}{034312} (\bibinfo{year}{2017}),
  \urlprefix\url{https://link.aps.org/doi/10.1103/PhysRevC.96.034312}.

\bibitem[{\citenamefont{Bender et~al.}(2003)\citenamefont{Bender, Heenen, and
  Reinhard}}]{Bender:2003jk}
\bibinfo{author}{\bibfnamefont{M.}~\bibnamefont{Bender}},
  \bibinfo{author}{\bibfnamefont{P.-H.} \bibnamefont{Heenen}},
  \bibnamefont{and} \bibinfo{author}{\bibfnamefont{P.-G.}
  \bibnamefont{Reinhard}}, \bibinfo{journal}{Rev. Mod. Phys.}
  \textbf{\bibinfo{volume}{75}}, \bibinfo{pages}{121} (\bibinfo{year}{2003}),
  \urlprefix\url{https://link.aps.org/doi/10.1103/RevModPhys.75.121}.

\bibitem[{\citenamefont{Egido}(2016)}]{Egido:2016bdz}
\bibinfo{author}{\bibfnamefont{J.~L.} \bibnamefont{Egido}},
  \bibinfo{journal}{Physica Scripta} \textbf{\bibinfo{volume}{91}},
  \bibinfo{pages}{073003} (\bibinfo{year}{2016}),
  \urlprefix\url{https://doi.org/10.1088/0031-8949/91/7/073003}.

\bibitem[{\citenamefont{Robledo et~al.}(2018)\citenamefont{Robledo,
  Rodr{\'{\i}}guez, and Rodr{\'{\i}}guez-Guzm{\'{a}}n}}]{Robledo:2018cdj}
\bibinfo{author}{\bibfnamefont{L.~M.} \bibnamefont{Robledo}},
  \bibinfo{author}{\bibfnamefont{T.~R.} \bibnamefont{Rodr{\'{\i}}guez}},
  \bibnamefont{and} \bibinfo{author}{\bibfnamefont{R.~R.}
  \bibnamefont{Rodr{\'{\i}}guez-Guzm{\'{a}}n}}, \bibinfo{journal}{Journal of
  Physics G: Nuclear and Particle Physics} \textbf{\bibinfo{volume}{46}},
  \bibinfo{pages}{013001} (\bibinfo{year}{2018}),
  \urlprefix\url{https://doi.org/10.1088/1361-6471/aadebd}.

\bibitem[{\citenamefont{Ripka}(1965)}]{ripk1}
\bibinfo{author}{\bibfnamefont{G.}~\bibnamefont{Ripka}},
  \emph{\bibinfo{title}{Lectures in Theoretical Physics: Volume VIII C}}
  (\bibinfo{publisher}{The University of Colorado Press},
  \bibinfo{year}{1965}).

\bibitem[{\citenamefont{Ripka}(1968)}]{Ripka1968}
\bibinfo{author}{\bibfnamefont{G.}~\bibnamefont{Ripka}},
  \emph{\bibinfo{title}{The Hartree-Fock Theory of Deformed Light Nuclei}}
  (\bibinfo{publisher}{Springer US}, \bibinfo{address}{Boston, MA},
  \bibinfo{year}{1968}), pp. \bibinfo{pages}{183--259}.

\bibitem[{\citenamefont{Schmid and Grummer}(1987)}]{VampirI}
\bibinfo{author}{\bibfnamefont{K.~W.} \bibnamefont{Schmid}} \bibnamefont{and}
  \bibinfo{author}{\bibfnamefont{F.}~\bibnamefont{Grummer}},
  \bibinfo{journal}{Reports on Progress in Physics}
  \textbf{\bibinfo{volume}{50}}, \bibinfo{pages}{731} (\bibinfo{year}{1987}),
  \urlprefix\url{https://doi.org/10.1088/0034-4885/50/6/003}.

\bibitem[{\citenamefont{Schmid}(2004)}]{VampirII}
\bibinfo{author}{\bibfnamefont{K.}~\bibnamefont{Schmid}},
  \bibinfo{journal}{Progress in Particle and Nuclear Physics}
  \textbf{\bibinfo{volume}{52}}, \bibinfo{pages}{565} (\bibinfo{year}{2004}),
  ISSN \bibinfo{issn}{0146-6410},
  \urlprefix\url{https://www.sciencedirect.com/science/article/pii/S0146641004000134}.

\bibitem[{\citenamefont{Otsuka et~al.}(1998)\citenamefont{Otsuka, Honma, and
  Mizusaki}}]{MCSM}
\bibinfo{author}{\bibfnamefont{T.}~\bibnamefont{Otsuka}},
  \bibinfo{author}{\bibfnamefont{M.}~\bibnamefont{Honma}}, \bibnamefont{and}
  \bibinfo{author}{\bibfnamefont{T.}~\bibnamefont{Mizusaki}},
  \bibinfo{journal}{Phys. Rev. Lett.} \textbf{\bibinfo{volume}{81}},
  \bibinfo{pages}{1588} (\bibinfo{year}{1998}),
  \urlprefix\url{https://link.aps.org/doi/10.1103/PhysRevLett.81.1588}.

\bibitem[{\citenamefont{Utsuno et~al.}(1999)\citenamefont{Utsuno, Otsuka,
  Mizusaki, and Honma}}]{MCSMII}
\bibinfo{author}{\bibfnamefont{Y.}~\bibnamefont{Utsuno}},
  \bibinfo{author}{\bibfnamefont{T.}~\bibnamefont{Otsuka}},
  \bibinfo{author}{\bibfnamefont{T.}~\bibnamefont{Mizusaki}}, \bibnamefont{and}
  \bibinfo{author}{\bibfnamefont{M.}~\bibnamefont{Honma}},
  \bibinfo{journal}{Phys. Rev. C} \textbf{\bibinfo{volume}{60}},
  \bibinfo{pages}{054315} (\bibinfo{year}{1999}),
  \urlprefix\url{https://link.aps.org/doi/10.1103/PhysRevC.60.054315}.

\bibitem[{\citenamefont{Tsunoda et~al.}(2014)\citenamefont{Tsunoda, Otsuka,
  Shimizu, Honma, and Utsuno}}]{MCSMIII}
\bibinfo{author}{\bibfnamefont{Y.}~\bibnamefont{Tsunoda}},
  \bibinfo{author}{\bibfnamefont{T.}~\bibnamefont{Otsuka}},
  \bibinfo{author}{\bibfnamefont{N.}~\bibnamefont{Shimizu}},
  \bibinfo{author}{\bibfnamefont{M.}~\bibnamefont{Honma}}, \bibnamefont{and}
  \bibinfo{author}{\bibfnamefont{Y.}~\bibnamefont{Utsuno}},
  \bibinfo{journal}{Phys. Rev. C} \textbf{\bibinfo{volume}{89}},
  \bibinfo{pages}{031301} (\bibinfo{year}{2014}),
  \urlprefix\url{https://link.aps.org/doi/10.1103/PhysRevC.89.031301}.

\bibitem[{\citenamefont{Gao and Horoi}(2009)}]{zch09}
\bibinfo{author}{\bibfnamefont{Z.-C.} \bibnamefont{Gao}} \bibnamefont{and}
  \bibinfo{author}{\bibfnamefont{M.}~\bibnamefont{Horoi}},
  \bibinfo{journal}{Phys. Rev. C} \textbf{\bibinfo{volume}{79}},
  \bibinfo{pages}{014311} (\bibinfo{year}{2009}),
  \urlprefix\url{https://link.aps.org/doi/10.1103/PhysRevC.79.014311}.

\bibitem[{\citenamefont{Hinohara and Kanada-En'yo}(2011)}]{hke11}
\bibinfo{author}{\bibfnamefont{N.}~\bibnamefont{Hinohara}} \bibnamefont{and}
  \bibinfo{author}{\bibfnamefont{Y.}~\bibnamefont{Kanada-En'yo}},
  \bibinfo{journal}{Phys. Rev. C} \textbf{\bibinfo{volume}{83}},
  \bibinfo{pages}{014321} (\bibinfo{year}{2011}),
  \urlprefix\url{https://link.aps.org/doi/10.1103/PhysRevC.83.014321}.

\bibitem[{\citenamefont{{Bally, B.}
  et~al.}(2021{\natexlab{a}})\citenamefont{{Bally, B.}, {S\'anchez-Fern\'andez,
  A.}, and {Rodr\'{\i}guez, T. R.}}}]{Taurus}
\bibinfo{author}{\bibnamefont{{Bally, B.}}},
  \bibinfo{author}{\bibnamefont{{S\'anchez-Fern\'andez, A.}}},
  \bibnamefont{and} \bibinfo{author}{\bibnamefont{{Rodr\'{\i}guez, T. R.}}},
  \bibinfo{journal}{Eur. Phys. J. A} \textbf{\bibinfo{volume}{57}},
  \bibinfo{pages}{69} (\bibinfo{year}{2021}{\natexlab{a}}),
  \urlprefix\url{https://doi.org/10.1140/epja/s10050-021-00369-z}.

\bibitem[{\citenamefont{S\'anchez-Fern\'andez
  et~al.}(2021)\citenamefont{S\'anchez-Fern\'andez, Bally, and
  Rodr\'{\i}guez}}]{Taurus-sd}
\bibinfo{author}{\bibfnamefont{A.}~\bibnamefont{S\'anchez-Fern\'andez}},
  \bibinfo{author}{\bibfnamefont{B.}~\bibnamefont{Bally}}, \bibnamefont{and}
  \bibinfo{author}{\bibfnamefont{T.~R.} \bibnamefont{Rodr\'{\i}guez}},
  \bibinfo{journal}{Phys. Rev. C} \textbf{\bibinfo{volume}{104}},
  \bibinfo{pages}{054306} (\bibinfo{year}{2021}),
  \urlprefix\url{https://link.aps.org/doi/10.1103/PhysRevC.104.054306}.

\bibitem[{\citenamefont{Bally et~al.}(2019)\citenamefont{Bally,
  S\'anchez-Fern\'andez, and Rodr\'{\i}guez}}]{Taurus-pf}
\bibinfo{author}{\bibfnamefont{B.}~\bibnamefont{Bally}},
  \bibinfo{author}{\bibfnamefont{A.}~\bibnamefont{S\'anchez-Fern\'andez}},
  \bibnamefont{and} \bibinfo{author}{\bibfnamefont{T.~R.}
  \bibnamefont{Rodr\'{\i}guez}}, \bibinfo{journal}{Phys. Rev. C}
  \textbf{\bibinfo{volume}{100}}, \bibinfo{pages}{044308}
  (\bibinfo{year}{2019}),
  \urlprefix\url{https://link.aps.org/doi/10.1103/PhysRevC.100.044308}.

\bibitem[{\citenamefont{Griffin and Wheeler}(1957)}]{GCM1957a}
\bibinfo{author}{\bibfnamefont{J.~J.} \bibnamefont{Griffin}} \bibnamefont{and}
  \bibinfo{author}{\bibfnamefont{J.~A.} \bibnamefont{Wheeler}},
  \bibinfo{journal}{Phys. Rev.} \textbf{\bibinfo{volume}{108}},
  \bibinfo{pages}{311} (\bibinfo{year}{1957}),
  \urlprefix\url{https://link.aps.org/doi/10.1103/PhysRev.108.311}.

\bibitem[{\citenamefont{Peierls and Yoccoz}(1957)}]{GCM1957b}
\bibinfo{author}{\bibfnamefont{R.~E.} \bibnamefont{Peierls}} \bibnamefont{and}
  \bibinfo{author}{\bibfnamefont{J.}~\bibnamefont{Yoccoz}},
  \bibinfo{journal}{Proceedings of the Physical Society. Section A}
  \textbf{\bibinfo{volume}{70}}, \bibinfo{pages}{381} (\bibinfo{year}{1957}),
  \urlprefix\url{https://doi.org/10.1088/0370-1298/70/5/309}.

\bibitem[{\citenamefont{Broeckhove and Deumens}(1979)}]{Deumens1979}
\bibinfo{author}{\bibfnamefont{J.}~\bibnamefont{Broeckhove}} \bibnamefont{and}
  \bibinfo{author}{\bibfnamefont{E.}~\bibnamefont{Deumens}},
  \bibinfo{journal}{Z. Phys. A} \textbf{\bibinfo{volume}{292}},
  \bibinfo{pages}{243} (\bibinfo{year}{1979}).

\bibitem[{\citenamefont{Caurier}(1975)}]{Caurier1975}
\bibinfo{author}{\bibfnamefont{E.}~\bibnamefont{Caurier}},
  \bibinfo{journal}{Proc. on GCM, BLG report} \textbf{\bibinfo{volume}{484}},
  \bibinfo{pages}{200} (\bibinfo{year}{1975}).

\bibitem[{\citenamefont{F.~Arickx and Leuven}(1981)}]{Arickx1981}
\bibinfo{author}{\bibfnamefont{E.~D.} \bibnamefont{F.~Arickx},
  \bibfnamefont{J.~Broeckhove}} \bibnamefont{and}
  \bibinfo{author}{\bibfnamefont{P.~V.} \bibnamefont{Leuven}},
  \bibinfo{journal}{J. Comp. Phys.} \textbf{\bibinfo{volume}{39}},
  \bibinfo{pages}{272} (\bibinfo{year}{1981}).

\bibitem[{\citenamefont{Johnson}(1988)}]{BroydenI}
\bibinfo{author}{\bibfnamefont{D.~D.} \bibnamefont{Johnson}},
  \bibinfo{journal}{Phys. Rev. B} \textbf{\bibinfo{volume}{38}},
  \bibinfo{pages}{12807} (\bibinfo{year}{1988}),
  \urlprefix\url{https://link.aps.org/doi/10.1103/PhysRevB.38.12807}.

\bibitem[{\citenamefont{Eyert}(1996)}]{BroydenII}
\bibinfo{author}{\bibfnamefont{V.}~\bibnamefont{Eyert}},
  \bibinfo{journal}{Journal of Computational Physics}
  \textbf{\bibinfo{volume}{124}}, \bibinfo{pages}{271} (\bibinfo{year}{1996}),
  ISSN \bibinfo{issn}{0021-9991},
  \urlprefix\url{https://www.sciencedirect.com/science/article/pii/S0021999196900595}.

\bibitem[{\citenamefont{Baran et~al.}(2008)\citenamefont{Baran, Bulgac, Forbes,
  Hagen, Nazarewicz, Schunck, and Stoitsov}}]{BroydenIII}
\bibinfo{author}{\bibfnamefont{A.}~\bibnamefont{Baran}},
  \bibinfo{author}{\bibfnamefont{A.}~\bibnamefont{Bulgac}},
  \bibinfo{author}{\bibfnamefont{M.~M.} \bibnamefont{Forbes}},
  \bibinfo{author}{\bibfnamefont{G.}~\bibnamefont{Hagen}},
  \bibinfo{author}{\bibfnamefont{W.}~\bibnamefont{Nazarewicz}},
  \bibinfo{author}{\bibfnamefont{N.}~\bibnamefont{Schunck}}, \bibnamefont{and}
  \bibinfo{author}{\bibfnamefont{M.~V.} \bibnamefont{Stoitsov}},
  \bibinfo{journal}{Phys. Rev. C} \textbf{\bibinfo{volume}{78}},
  \bibinfo{pages}{014318} (\bibinfo{year}{2008}),
  \urlprefix\url{https://link.aps.org/doi/10.1103/PhysRevC.78.014318}.

\bibitem[{\citenamefont{{Staszczak, A.} et~al.}(2010)\citenamefont{{Staszczak,
  A.}, {Stoitsov, M.}, {Baran, A.}, and {Nazarewicz, W.}}}]{augLag}
\bibinfo{author}{\bibnamefont{{Staszczak, A.}}},
  \bibinfo{author}{\bibnamefont{{Stoitsov, M.}}},
  \bibinfo{author}{\bibnamefont{{Baran, A.}}}, \bibnamefont{and}
  \bibinfo{author}{\bibnamefont{{Nazarewicz, W.}}}, \bibinfo{journal}{Eur.
  Phys. J. A} \textbf{\bibinfo{volume}{46}}, \bibinfo{pages}{85}
  (\bibinfo{year}{2010}),
  \urlprefix\url{https://doi.org/10.1140/epja/i2010-11018-9}.

\bibitem[{\citenamefont{Lamme and Boeker}(1968)}]{projI}
\bibinfo{author}{\bibfnamefont{H.}~\bibnamefont{Lamme}} \bibnamefont{and}
  \bibinfo{author}{\bibfnamefont{E.}~\bibnamefont{Boeker}},
  \bibinfo{journal}{Nuclear Physics A} \textbf{\bibinfo{volume}{111}},
  \bibinfo{pages}{492} (\bibinfo{year}{1968}), ISSN \bibinfo{issn}{0375-9474},
  \urlprefix\url{https://www.sciencedirect.com/science/article/pii/0375947468902352}.

\bibitem[{\citenamefont{Messiah}(1965)}]{WignerD}
\bibinfo{author}{\bibfnamefont{A.}~\bibnamefont{Messiah}},
  \emph{\bibinfo{title}{Quantum mechanics}} (\bibinfo{publisher}{North-Holland
  Publishing Co., Amsterdam}, \bibinfo{year}{1965}).

\bibitem[{\citenamefont{Bertsch et~al.}(2009)\citenamefont{Bertsch,
  Dobaczewski, Nazarewicz, and Pei}}]{crankingI}
\bibinfo{author}{\bibfnamefont{G.}~\bibnamefont{Bertsch}},
  \bibinfo{author}{\bibfnamefont{J.}~\bibnamefont{Dobaczewski}},
  \bibinfo{author}{\bibfnamefont{W.}~\bibnamefont{Nazarewicz}},
  \bibnamefont{and} \bibinfo{author}{\bibfnamefont{J.}~\bibnamefont{Pei}},
  \bibinfo{journal}{Phys. Rev. A} \textbf{\bibinfo{volume}{79}},
  \bibinfo{pages}{043602} (\bibinfo{year}{2009}),
  \urlprefix\url{https://link.aps.org/doi/10.1103/PhysRevA.79.043602}.

\bibitem[{\citenamefont{Kasuya and Yoshida}(2021)}]{crankingII}
\bibinfo{author}{\bibfnamefont{H.}~\bibnamefont{Kasuya}} \bibnamefont{and}
  \bibinfo{author}{\bibfnamefont{K.}~\bibnamefont{Yoshida}},
  \bibinfo{journal}{Prog. Theo. Exp. Phys.} \textbf{\bibinfo{volume}{2021}}
  (\bibinfo{year}{2021}).

\bibitem[{\citenamefont{Watt}(1972)}]{Watt1972}
\bibinfo{author}{\bibfnamefont{A.}~\bibnamefont{Watt}}, \bibinfo{journal}{J.
  Phys. A: Gen. Phys.} \textbf{\bibinfo{volume}{5}} (\bibinfo{year}{1972}).

\bibitem[{\citenamefont{Rodr\'{\i}guez and Egido}(2010)}]{GCMtriaxI}
\bibinfo{author}{\bibfnamefont{T.~R.} \bibnamefont{Rodr\'{\i}guez}}
  \bibnamefont{and} \bibinfo{author}{\bibfnamefont{J.~L.} \bibnamefont{Egido}},
  \bibinfo{journal}{Phys. Rev. C} \textbf{\bibinfo{volume}{81}},
  \bibinfo{pages}{064323} (\bibinfo{year}{2010}),
  \urlprefix\url{https://link.aps.org/doi/10.1103/PhysRevC.81.064323}.

\bibitem[{\citenamefont{Bender and Heenen}(2008)}]{GCMtriaxII}
\bibinfo{author}{\bibfnamefont{M.}~\bibnamefont{Bender}} \bibnamefont{and}
  \bibinfo{author}{\bibfnamefont{P.-H.} \bibnamefont{Heenen}},
  \bibinfo{journal}{Phys. Rev. C} \textbf{\bibinfo{volume}{78}},
  \bibinfo{pages}{024309} (\bibinfo{year}{2008}),
  \urlprefix\url{https://link.aps.org/doi/10.1103/PhysRevC.78.024309}.

\bibitem[{\citenamefont{Burzy\ifmmode~\acute{n}\else \'{n}\fi{}ski and
  Dobaczewski}(1995)}]{GCMtriaxIII}
\bibinfo{author}{\bibfnamefont{K.}~\bibnamefont{Burzy\ifmmode~\acute{n}\else
  \'{n}\fi{}ski}} \bibnamefont{and}
  \bibinfo{author}{\bibfnamefont{J.}~\bibnamefont{Dobaczewski}},
  \bibinfo{journal}{Phys. Rev. C} \textbf{\bibinfo{volume}{51}},
  \bibinfo{pages}{1825} (\bibinfo{year}{1995}),
  \urlprefix\url{https://link.aps.org/doi/10.1103/PhysRevC.51.1825}.

\bibitem[{\citenamefont{Ring and Schuck}(1980)}]{Ring1980}
\bibinfo{author}{\bibfnamefont{P.}~\bibnamefont{Ring}} \bibnamefont{and}
  \bibinfo{author}{\bibfnamefont{P.}~\bibnamefont{Schuck}},
  \emph{\bibinfo{title}{The Nuclear Many-Body Problem}}
  (\bibinfo{publisher}{Springer-Verlag, Heidelberg, 1980},
  \bibinfo{year}{1980}).

\bibitem[{\citenamefont{Hill and Wheeler}(1953)}]{betagamma}
\bibinfo{author}{\bibfnamefont{D.~L.} \bibnamefont{Hill}} \bibnamefont{and}
  \bibinfo{author}{\bibfnamefont{J.~A.} \bibnamefont{Wheeler}},
  \bibinfo{journal}{Phys. Rev.} \textbf{\bibinfo{volume}{89}},
  \bibinfo{pages}{1102} (\bibinfo{year}{1953}),
  \urlprefix\url{https://link.aps.org/doi/10.1103/PhysRev.89.1102}.

\bibitem[{\citenamefont{Brussard and Glaudemans}(1977)}]{Brussard1977}
\bibinfo{author}{\bibfnamefont{P.~J.} \bibnamefont{Brussard}} \bibnamefont{and}
  \bibinfo{author}{\bibfnamefont{P.~W.~M.} \bibnamefont{Glaudemans}},
  \emph{\bibinfo{title}{Shell Model Applications in Nuclear Spectroscopy}}
  (\bibinfo{publisher}{North-Holland Publishing Co., Amsterdam .New York.
  Oxford}, \bibinfo{year}{1977}).

\bibitem[{\citenamefont{Suhonen}(2007)}]{js07}
\bibinfo{author}{\bibfnamefont{J.}~\bibnamefont{Suhonen}},
  \emph{\bibinfo{title}{From Nucleons to Nucleus, Concepts of Microscopic
  Nuclear Theory}} (\bibinfo{publisher}{Springer-Verlag Berlin Heidelberg},
  \bibinfo{year}{2007}).

\bibitem[{\citenamefont{Zuker et~al.}(2015)\citenamefont{Zuker, Poves, Nowacki,
  and Lenzi}}]{NilssonSU3}
\bibinfo{author}{\bibfnamefont{A.~P.} \bibnamefont{Zuker}},
  \bibinfo{author}{\bibfnamefont{A.}~\bibnamefont{Poves}},
  \bibinfo{author}{\bibfnamefont{F.}~\bibnamefont{Nowacki}}, \bibnamefont{and}
  \bibinfo{author}{\bibfnamefont{S.~M.} \bibnamefont{Lenzi}},
  \bibinfo{journal}{Phys. Rev. C} \textbf{\bibinfo{volume}{92}},
  \bibinfo{pages}{024320} (\bibinfo{year}{2015}),
  \urlprefix\url{https://link.aps.org/doi/10.1103/PhysRevC.92.024320}.

\bibitem[{\citenamefont{Coraggio and Itaco}(2020)}]{Coraggio:2020cmw}
\bibinfo{author}{\bibfnamefont{L.}~\bibnamefont{Coraggio}} \bibnamefont{and}
  \bibinfo{author}{\bibfnamefont{N.}~\bibnamefont{Itaco}},
  \bibinfo{journal}{Frontiers in Physics} \textbf{\bibinfo{volume}{8}}
  (\bibinfo{year}{2020}), ISSN \bibinfo{issn}{2296-424X},
  \urlprefix\url{https://www.frontiersin.org/article/10.3389/fphy.2020.00345}.

\bibitem[{\citenamefont{Dufour and Zuker}(1996)}]{hmulti}
\bibinfo{author}{\bibfnamefont{M.}~\bibnamefont{Dufour}} \bibnamefont{and}
  \bibinfo{author}{\bibfnamefont{A.~P.} \bibnamefont{Zuker}},
  \bibinfo{journal}{Phys. Rev. C} \textbf{\bibinfo{volume}{54}},
  \bibinfo{pages}{1641} (\bibinfo{year}{1996}),
  \urlprefix\url{https://link.aps.org/doi/10.1103/PhysRevC.54.1641}.

\bibitem[{\citenamefont{Brown and Richter}(2006)}]{Brown:USDB}
\bibinfo{author}{\bibfnamefont{B.~A.} \bibnamefont{Brown}} \bibnamefont{and}
  \bibinfo{author}{\bibfnamefont{W.~A.} \bibnamefont{Richter}},
  \bibinfo{journal}{Phys. Rev. C} \textbf{\bibinfo{volume}{74}},
  \bibinfo{pages}{034315} (\bibinfo{year}{2006}),
  \urlprefix\url{https://link.aps.org/doi/10.1103/PhysRevC.74.034315}.

\bibitem[{\citenamefont{Yao et~al.}(2009)\citenamefont{Yao, Meng, Ring, and
  Arteaga}}]{Mg24Yao}
\bibinfo{author}{\bibfnamefont{J.~M.} \bibnamefont{Yao}},
  \bibinfo{author}{\bibfnamefont{J.}~\bibnamefont{Meng}},
  \bibinfo{author}{\bibfnamefont{P.}~\bibnamefont{Ring}}, \bibnamefont{and}
  \bibinfo{author}{\bibfnamefont{D.~P.} \bibnamefont{Arteaga}},
  \bibinfo{journal}{Phys. Rev. C} \textbf{\bibinfo{volume}{79}},
  \bibinfo{pages}{044312} (\bibinfo{year}{2009}),
  \urlprefix\url{https://link.aps.org/doi/10.1103/PhysRevC.79.044312}.

\bibitem[{\citenamefont{Yao et~al.}(2010)\citenamefont{Yao, Meng, Ring, and
  Vretenar}}]{Mg24YaoII}
\bibinfo{author}{\bibfnamefont{J.~M.} \bibnamefont{Yao}},
  \bibinfo{author}{\bibfnamefont{J.}~\bibnamefont{Meng}},
  \bibinfo{author}{\bibfnamefont{P.}~\bibnamefont{Ring}}, \bibnamefont{and}
  \bibinfo{author}{\bibfnamefont{D.}~\bibnamefont{Vretenar}},
  \bibinfo{journal}{Phys. Rev. C} \textbf{\bibinfo{volume}{81}},
  \bibinfo{pages}{044311} (\bibinfo{year}{2010}),
  \urlprefix\url{https://link.aps.org/doi/10.1103/PhysRevC.81.044311}.

\bibitem[{\citenamefont{Borrajo et~al.}(2015)\citenamefont{Borrajo, Rodríguez,
  and {Luis Egido}}}]{Tomas2015cranking}
\bibinfo{author}{\bibfnamefont{M.}~\bibnamefont{Borrajo}},
  \bibinfo{author}{\bibfnamefont{T.~R.} \bibnamefont{Rodríguez}},
  \bibnamefont{and} \bibinfo{author}{\bibfnamefont{J.}~\bibnamefont{{Luis
  Egido}}}, \bibinfo{journal}{Physics Letters B}
  \textbf{\bibinfo{volume}{746}}, \bibinfo{pages}{341} (\bibinfo{year}{2015}),
  ISSN \bibinfo{issn}{0370-2693},
  \urlprefix\url{https://www.sciencedirect.com/science/article/pii/S0370269315003676}.

\bibitem[{\citenamefont{Honma et~al.}(1996)\citenamefont{Honma, Mizusaki, and
  Otsuka}}]{MCSM1996Mg24}
\bibinfo{author}{\bibfnamefont{M.}~\bibnamefont{Honma}},
  \bibinfo{author}{\bibfnamefont{T.}~\bibnamefont{Mizusaki}}, \bibnamefont{and}
  \bibinfo{author}{\bibfnamefont{T.}~\bibnamefont{Otsuka}},
  \bibinfo{journal}{Phys. Rev. Lett.} \textbf{\bibinfo{volume}{77}},
  \bibinfo{pages}{3315} (\bibinfo{year}{1996}),
  \urlprefix\url{https://link.aps.org/doi/10.1103/PhysRevLett.77.3315}.

\bibitem[{\citenamefont{Wang et~al.}(2018)\citenamefont{Wang, Gao, Ma, and
  Chen}}]{Gao2018}
\bibinfo{author}{\bibfnamefont{J.-Q.} \bibnamefont{Wang}},
  \bibinfo{author}{\bibfnamefont{Z.-C.} \bibnamefont{Gao}},
  \bibinfo{author}{\bibfnamefont{Y.-J.} \bibnamefont{Ma}}, \bibnamefont{and}
  \bibinfo{author}{\bibfnamefont{Y.~S.} \bibnamefont{Chen}},
  \bibinfo{journal}{Phys. Rev. C} \textbf{\bibinfo{volume}{98}},
  \bibinfo{pages}{021301} (\bibinfo{year}{2018}),
  \urlprefix\url{https://link.aps.org/doi/10.1103/PhysRevC.98.021301}.

\bibitem[{\citenamefont{{Bally, B.}
  et~al.}(2021{\natexlab{b}})\citenamefont{{Bally, B.}, {S\'anchez-Fern\'andez,
  A.}, and {Rodr\'{\i}guez, T. R.}}}]{Taurus-erratum}
\bibinfo{author}{\bibnamefont{{Bally, B.}}},
  \bibinfo{author}{\bibnamefont{{S\'anchez-Fern\'andez, A.}}},
  \bibnamefont{and} \bibinfo{author}{\bibnamefont{{Rodr\'{\i}guez, T. R.}}},
  \bibinfo{journal}{Eur. Phys. J. A} \textbf{\bibinfo{volume}{57}},
  \bibinfo{pages}{124} (\bibinfo{year}{2021}{\natexlab{b}}),
  \urlprefix\url{https://doi.org/10.1140/epja/s10050-021-00406-x}.

\bibitem[{\citenamefont{Caurier et~al.}(2003)\citenamefont{Caurier, Rejmund,
  and Grawe}}]{N=126}
\bibinfo{author}{\bibfnamefont{E.}~\bibnamefont{Caurier}},
  \bibinfo{author}{\bibfnamefont{M.}~\bibnamefont{Rejmund}}, \bibnamefont{and}
  \bibinfo{author}{\bibfnamefont{H.}~\bibnamefont{Grawe}},
  \bibinfo{journal}{Phys. Rev. C} \textbf{\bibinfo{volume}{67}},
  \bibinfo{pages}{054310} (\bibinfo{year}{2003}),
  \urlprefix\url{https://link.aps.org/doi/10.1103/PhysRevC.67.054310}.

\bibitem[{\citenamefont{Hauschild et~al.}(2001)\citenamefont{Hauschild,
  Rejmund, Grawe, Caurier, Nowacki, Becker, Le~Coz, Korten, D\"oring, G\'orska
  et~al.}}]{Th216}
\bibinfo{author}{\bibfnamefont{K.}~\bibnamefont{Hauschild}},
  \bibinfo{author}{\bibfnamefont{M.}~\bibnamefont{Rejmund}},
  \bibinfo{author}{\bibfnamefont{H.}~\bibnamefont{Grawe}},
  \bibinfo{author}{\bibfnamefont{E.}~\bibnamefont{Caurier}},
  \bibinfo{author}{\bibfnamefont{F.}~\bibnamefont{Nowacki}},
  \bibinfo{author}{\bibfnamefont{F.}~\bibnamefont{Becker}},
  \bibinfo{author}{\bibfnamefont{Y.}~\bibnamefont{Le~Coz}},
  \bibinfo{author}{\bibfnamefont{W.}~\bibnamefont{Korten}},
  \bibinfo{author}{\bibfnamefont{J.}~\bibnamefont{D\"oring}},
  \bibinfo{author}{\bibfnamefont{M.}~\bibnamefont{G\'orska}},
  \bibnamefont{et~al.}, \bibinfo{journal}{Phys. Rev. Lett.}
  \textbf{\bibinfo{volume}{87}}, \bibinfo{pages}{072501}
  (\bibinfo{year}{2001}),
  \urlprefix\url{https://link.aps.org/doi/10.1103/PhysRevLett.87.072501}.

\bibitem[{\citenamefont{Herzberg et~al.}(2006)\citenamefont{Herzberg,
  Greenlees, and Butler}}]{No254-Nature}
\bibinfo{author}{\bibfnamefont{R.}~\bibnamefont{Herzberg}},
  \bibinfo{author}{\bibfnamefont{P.}~\bibnamefont{Greenlees}},
  \bibnamefont{and} \bibinfo{author}{\bibfnamefont{P.}~\bibnamefont{Butler}},
  \bibinfo{journal}{Nature} \textbf{\bibinfo{volume}{442}},
  \bibinfo{pages}{896–899} (\bibinfo{year}{2006}).

\bibitem[{\citenamefont{Clark et~al.}(2010)\citenamefont{Clark, Gregorich,
  Berryman, Ali, Allmond, Beausang, Cromaz, Deleplanque, Dragojević, Dvorak
  et~al.}}]{No254-PLB}
\bibinfo{author}{\bibfnamefont{R.}~\bibnamefont{Clark}},
  \bibinfo{author}{\bibfnamefont{K.}~\bibnamefont{Gregorich}},
  \bibinfo{author}{\bibfnamefont{J.}~\bibnamefont{Berryman}},
  \bibinfo{author}{\bibfnamefont{M.}~\bibnamefont{Ali}},
  \bibinfo{author}{\bibfnamefont{J.}~\bibnamefont{Allmond}},
  \bibinfo{author}{\bibfnamefont{C.}~\bibnamefont{Beausang}},
  \bibinfo{author}{\bibfnamefont{M.}~\bibnamefont{Cromaz}},
  \bibinfo{author}{\bibfnamefont{M.}~\bibnamefont{Deleplanque}},
  \bibinfo{author}{\bibfnamefont{I.}~\bibnamefont{Dragojević}},
  \bibinfo{author}{\bibfnamefont{J.}~\bibnamefont{Dvorak}},
  \bibnamefont{et~al.}, \bibinfo{journal}{Physics Letters B}
  \textbf{\bibinfo{volume}{690}}, \bibinfo{pages}{19} (\bibinfo{year}{2010}),
  ISSN \bibinfo{issn}{0370-2693},
  \urlprefix\url{https://www.sciencedirect.com/science/article/pii/S0370269310005757}.

\bibitem[{\citenamefont{Calinescu et~al.}(2021)\citenamefont{Calinescu, Sorlin,
  Matea, Carstoiu, Dao, Nowacki, de~Angelis, Astabatyan, Bagchi, Borcea
  et~al.}}]{Zn70}
\bibinfo{author}{\bibfnamefont{S.}~\bibnamefont{Calinescu}},
  \bibinfo{author}{\bibfnamefont{O.}~\bibnamefont{Sorlin}},
  \bibinfo{author}{\bibfnamefont{I.}~\bibnamefont{Matea}},
  \bibinfo{author}{\bibfnamefont{F.}~\bibnamefont{Carstoiu}},
  \bibinfo{author}{\bibfnamefont{D.~D.} \bibnamefont{Dao}},
  \bibinfo{author}{\bibfnamefont{F.}~\bibnamefont{Nowacki}},
  \bibinfo{author}{\bibfnamefont{G.}~\bibnamefont{de~Angelis}},
  \bibinfo{author}{\bibfnamefont{R.}~\bibnamefont{Astabatyan}},
  \bibinfo{author}{\bibfnamefont{S.}~\bibnamefont{Bagchi}},
  \bibinfo{author}{\bibfnamefont{C.}~\bibnamefont{Borcea}},
  \bibnamefont{et~al.}, \bibinfo{journal}{Phys. Rev. C}
  \textbf{\bibinfo{volume}{104}}, \bibinfo{pages}{034318}
  (\bibinfo{year}{2021}),
  \urlprefix\url{https://link.aps.org/doi/10.1103/PhysRevC.104.034318}.

\bibitem[{\citenamefont{Rezynkina et~al.}(2022)}]{Arsenics}
\bibinfo{author}{\bibfnamefont{K.}~\bibnamefont{Rezynkina}}
  \bibnamefont{et~al.}, \bibinfo{journal}{to be submitted to Phys. Rev.}
  (\bibinfo{year}{2022}).

\bibitem[{\citenamefont{Reygadas et~al.}(2022)}]{Bromes}
\bibinfo{author}{\bibfnamefont{D.}~\bibnamefont{Reygadas}}
  \bibnamefont{et~al.}, \bibinfo{journal}{to be submitted to Phys. Rev.}
  (\bibinfo{year}{2022}).

\bibitem[{\citenamefont{Rocchini et~al.}(2022)}]{Zn74}
\bibinfo{author}{\bibfnamefont{M.}~\bibnamefont{Rocchini}}
  \bibnamefont{et~al.}, \bibinfo{journal}{to be submitted to Phys. Rev.}
  (\bibinfo{year}{2022}).

\end{thebibliography}

\end{document}